\documentclass[11pt]{article}
\usepackage{graphicx}
\usepackage{amsmath}
\usepackage{amssymb}
\usepackage{latexsym}
\usepackage{bm}
\usepackage{overpic}
\usepackage{cases}
\usepackage{slashbox}
\usepackage[dvips]{color}
\newcommand{\ds}{\displaystyle}
\newtheorem{theorem}{\bf Theorem}
\newtheorem{lemma}{\bf Lemma}
\newtheorem{example}{\bf Example}
\begin{document}
\begin{center}
{\LARGE\bf On the Search Algorithm for the Output Distribution\\[2mm]
that Achieves the Channel Capacity}\\[7mm]
{\large Kenji~Nakagawa$^\ast$, Kohei~Watabe$^\ast$, Takuto~Sabu$^\ast$}
\end{center}
\footnote[0]{This paper was presented in part at SITA2014, STW2015, SITA2015. $^\ast$Department of Electrical and Electronics and Information Engineering, Nagaoka University of Technology, Nagaoka, Niigata 940-2188, Japan, e-mail:nakagawa@nagaokaut.ac.jp}
\begin{abstract}
We consider a search algorithm for the output distribution that achieves the channel capacity of a discrete memoryless channel. We will propose an algorithm by iterated projections of an output distribution onto affine subspaces in the set of output distributions. The problem of channel capacity has a similar geometric structure as that of smallest enclosing circle for a finite number of points in the Euclidean space. The metric in the Euclidean space is the Euclidean distance and the metric in the space of output distributions is the Kullback-Leibler divergence. We consider these two problems based on Amari's $\alpha$-geometry \cite{ama}. Then, we first consider the smallest enclosing circle in the Euclidean space and develop an algorithm to find the center of the smallest enclosing circle. Based on the investigation, we will apply the obtained algorithm to the problem of channel capacity.
\end{abstract}



\baselineskip 6mm

\section{Introduction}
The channel capacity $C$ of a discrete memoryless channel is defined as the maximum of the mutual information. $C$ is also formulated as the solution of a $\min\max$ problem concerned with the Kullback-Leibler divergence \cite{csi}, \cite{nak}. If we replace the Kullback-Leibler divergence with the Euclidean distance, a similar problem in the Euclidean space is obtained. That is the problem of smallest enclosing circle for a finite number of points. In this paper, we will investigate the problem of smallest enclosing circle in the Euclidean space geometrically, and develop an algorithm to compute the solution to the $\min\max$ problem of the Euclidean distance. Then, the resulting algorithm will be applied to the $\min\max$ problem of channel capacity to make an algorithm for calculating the output distribution that achieves the channel capacity. The reason for taking such an approach is because the Euclidean geometry is familiar to us, so it may be easier to make new geometric algorithms.

As mentioned above, the problem of channel capacity is described by an optimization problem concerned with the output probability distributions. The geometry on the set of output probability distributions is the information geometry \cite{ama}. The Euclidean geometry and the information geometry can be considered in a unified manner from the viewpoint of $\alpha$-geometry by Amari \cite{ama}. Therefore, the solution algorithm for the problem of smallest enclosing circle can be applied to the problem of channel capacity through this geometric similarity. For that purpose, it is necessary to use only the applicable properties to the problem of channel capacity. If it is achieved, then an algorithm obtained in the smallest enclosing circle can be transplanted almost automatically to the channel capacity. In this paper, we will actually use only barycentric coordinate, inner product, Pythagorean theorem, and projection onto affine subspaces, as common properties in both geometries, to develop a computation algorithm. The algorithm is called a ``projection algorithm.'' 

In this paper, first we consider the algorithm for calculating the center of the smallest enclosing circle for a finite number of points in general position in the Euclidean space. Then, similarly, in the case that the row vectors of the channel matrix are in general position, we consider the algorithm for calculating the output distribution that achieves the channel capacity. We will show that the both problems are solved by common geometric properties. Further, based on the above investigation, we consider the case that the finite number of points and the row vectors are not necessarily in general position. Then, finally we propose heuristic search algorithms and perform the proposed algorithm for randomly generated placements of points and row vectors. We evaluate the percentage that correct solutions are obtained by our heuristic algorithm.
\subsection{Related Works}
There are roughly two categories of calculation methods for the channel capacity, one is solving equations due to Muroga \cite{mur} and the other is a sequential calculation method due to Arimoto \cite {ari}. In Muroga \cite{mur}, the input probability distribution that achieves the channel capacity is obtained by solving directly the equations derived from the Lagrange multiplier method. In this case, we should take care of the inequality condition that the probability takes a non-negative value. Simply solving the equations by ignoring this inequality condition may yield a solution of a negative probability. Of course this is not a correct solution. The inequality condition makes the problem difficult, however, we can say this difficulty makes the problem attractive. In Muroga \cite{mur}, if a solution includes a negative probability, one of the input symbols should be removed and solving equations should be continued repeatedly until all the probabilities become non-negative. Therefore, if all the possibilities are exhausted, the total number of equations to be solved is exponential. 

In Arimoto's sequential approximation method \cite{ari}, the channel capacity of an arbitrary channel matrix is calculated numerically by a recurrence formula. This method does not yield a negative probability, but corresponding to it, if some input probability becomes 0, an exceptional treatment is required in the calculation of a reverse channel. Both in Muroga \cite{mur} and Arimoto \cite{ari}, it is important to determine which input symbol has a positive probability. That is a main subject of this paper.
\subsection{Channel matrix and channel capacity}
Let us consider a discrete memoryless channel $X\rightarrow Y$ with input source $X$ and output source $Y$. Denote by $\{x_1,\cdots,x_m\}$ the input alphabet and $\{y_1,\cdots,y_n\}$ the output alphabet. The conditional probability $P^i_j$ that $y_j$ is received when $x_i$ was transmitted is denoted by
\begin{align*}
P^i_j=P(Y=y_j|X=x_i),\,i=1,\cdots,m, j=1,\cdots,n,
\end{align*}
and the row vector $P^i$ is defined by
\begin{align}
P^i=(P^i_1,\cdots,P^i_n),\,i=1,\cdots,m.
\end{align}
The channel matrix $\Phi$ is defined by
\begin{align}
\label{eqn:thechannelmatrix}
\Phi=\left(\begin{array}{c}
P^1\\
\vdots\\
P^m
\end{array}\right)
=\left(\begin{array}{ccc}
P^1_1 & \cdots & P^1_n\\
\vdots & & \vdots\\
P^m_1 & \cdots & P^m_n
\end{array}\right).
\end{align}
The set $\bar\Delta^m$ of all input probability distributions on the input alphabet $\{x_1,\cdots,x_m\}$ is defined by 
\begin{align*}
\bar\Delta^m=\{\bm\lambda=(\lambda_1,\cdots,\lambda_m)|\lambda_i\geq0,i=1,\cdots,m,\sum_{i=1}^m\lambda_i=1\}.
\end{align*}
Similarly, the set $\bar\Delta^n$ of all output probability distributions on the output alphabet $\{y_1,\cdots,y_n\}$ is defined by 
\begin{align*}
\bar\Delta^n=\{Q=(Q_1,\cdots,Q_n)|Q_j\geq0,j=1,\cdots,n,\sum_{j=1}^nQ_j=1\}.
\end{align*}
The output distribution $Q\in\bar\Delta^n$ corresponding to the input distribution $\bm\lambda\in\bar\Delta^m$ is denoted by $Q=\bm\lambda\Phi$, 
i.e., $Q_j=\sum_{i=1}^m\lambda_iP^i_j,\,j=1,\cdots,n$, and the mutual information $I(\bm\lambda,\Phi)$ is defined by
\begin{align}
I(\bm\lambda,\Phi)=\sum_{i=1}^m\sum_{j=1}^n\lambda_iP^i_j\log\ds\frac{P^i_j}{Q_j}.
\end{align}
Then, the channel capacity $C$ is defined by
\begin{align}
\label{eqn:Cdefinition}
C=\max_{\bm\lambda\in\bar\Delta^m}I(\bm\lambda,\Phi).
\end{align}
For two output distributions $Q=(Q_1,\cdots,Q_n)$, $Q'=(Q'_1,\cdots,Q'_n)\in\bar\Delta^n$, the Kullback-Leibler divergence $D(Q\|Q')$ is defined by
\begin{align}
D(Q\|Q')\equiv\sum_{j=1}^nQ_j\log\ds\frac{Q_j}{Q'_j},
\end{align}
see \cite{csi}. The channel capacity $C$ is also formulated by the Kullback-Leibler divergence as follows \cite{csi}:
\begin{align}
C=\min_{Q\in\bar{\Delta}^n}\max_{1\leq i\leq m}D(P^i\|Q).\label{eqn:minmax2}
\end{align}
For some channel matrix, the input distribution $\bm\lambda$ that achieves (\ref{eqn:Cdefinition}) is not unique, but the output distribution $Q$ that achieves (\ref{eqn:minmax2}) is unique for any channel matrix \cite{csi}. By virtue of the uniqueness, it is easy to consider the method of calculating the channel capacity $C$ based on (\ref{eqn:minmax2}) using geometric properties of the Kullback-Leibler divergence. 

On the other hand, in order to prove that the resulting output distribution actually achieves $C$, we will use the convex optimization (\ref{eqn:Cdefinition}) rather than the geometrical consideration by (\ref{eqn:minmax2}). Concerning (\ref{eqn:Cdefinition}), the following Kuhn-Tucker condition holds \cite{gal}.

{\bf Theorem:} (Kuhn-Tucker condition for the problem of channel capacity) A necessary and sufficient condition for an input distribution $\bm\lambda^\ast=(\lambda^\ast_1,\cdots,\lambda^\ast_m)\in\bar\Delta^m$ to achieve the channel capacity $C$ is that there exists a value $C_0$ with
\begin{align}
\label{eqn:KT2}
D(P^i\|\bm\lambda^\ast\Phi)\left\{\begin{array}{ll}=C_0, & {\mbox{\rm for}}\ i\ {\mbox{\rm with}}\ \lambda^\ast_i>0,\\
\leq C_0, & {\mbox{\rm for}}\ i\ {\mbox{\rm with}}\ \lambda^\ast_i=0.
\end{array}\right.
\end{align}
Then, $C_0$ is equal to $C$.

\subsection{Smallest enclosing circle}
Now, replacing $D(P^i\|Q)$ in (\ref{eqn:minmax2}) with the Euclidean distance, we can consider a similar problem in $\mathbb{R}^n$. That is, for $P^1,\cdots,P^m\in\mathbb{R}^n$, let us consider
\begin{align}
\label{eqn:minimaxd}
\min_{Q\in\mathbb{R}^n}\max_{1\leq i\leq m}d(P^i,Q),
\end{align}
where $d(P^i,Q)$ denotes the Euclidean distance between the points $P^i$ and $Q$ in $\mathbb{R}^n$. This is the problem of smallest enclosing circle for the set of points $\{P^1,\cdots,P^m\}$. The purpose of this paper is to study the problem of smallest enclosing circle geometrically and obtain a search algorithm for the optimal solution. Then, through the similarity of (\ref{eqn:minmax2}) and (\ref{eqn:minimaxd}), we will apply the resulting algorithm to obtain a search algorithm for the output distribution that achieves the channel capacity.

(\ref{eqn:minmax2}) and (\ref{eqn:minimaxd}) not only resemble formally, but have common geometric structures from the view point of Amari's $\alpha$-geometry  \cite{ama}. Therefore, if we can develop an algorithm and prove its validity by using only the common properties to both geometries, then an algorithm obtained in one problem can be applied to the other problem almost automatically. In fact, it will be apparent that is so in this paper.

Then, first, let us consider the problem of smallest enclosing circle in the Euclidean space.
\section{Problem of smallest enclosing circle in the Euclidean space}
\label{sec:minimumenclosingcircle}
Consider a finite number of points $P^1,\cdots,P^m$ in the $n$ dimensional Euclidean space $\mathbb{R}^n$. The smallest sphere in $\mathbb{R}^n$ that includes these points in its inside or on the boundary is called the {\it smallest enclosing circle}, and is represented by $\Gamma(P^1,\cdots,P^m)$. The smallest enclosing circle $\Gamma(P^1,\cdots,P^m)$ is formulated by (\ref{eqn:minimaxd}). The $Q=Q^\ast$ that achieves (\ref{eqn:minimaxd}) is the center of $\Gamma(P^1,\cdots,P^m)$ and $d^\ast=\max_{1\leq i\leq m}d(P^i,Q^\ast)$ is its radius.
\subsection{Equidistant point and projection}
\label{subsec:eikakudonkaku}
As a simplest example, let us consider the center $Q^\ast$ and the radius $d^\ast$ of the smallest enclosing circle $\Gamma=\Gamma(P^1,P^2,P^3)$ for three points $P^1,P^2,P^3$ in $\mathbb{R}^2$. We will investigate separately in cases that $\triangle P^1P^2P^3$ is an acute triangle and obtuse triangle.

\begin{figure}[t]
\begin{center}
\begin{overpic}[width=6.6cm]{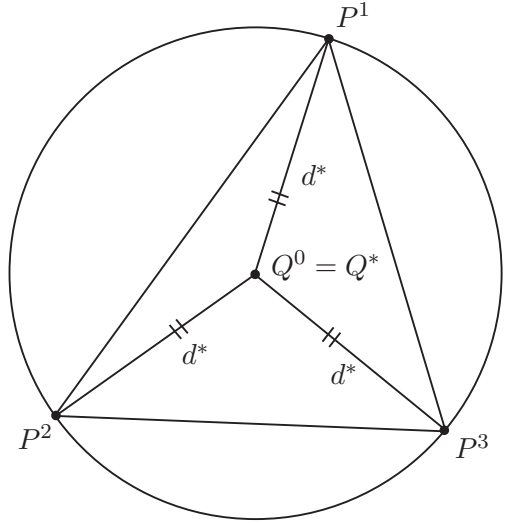}
\put(66,100){$P^1$}
\put(2,15){$P^2$}
\put(90,13){$P^3$}
\put(53,50){$Q^0=Q^\ast$}
\put(59,68){$d^\ast$}
\put(35,32){$d^\ast$}
\put(65,28){$d^\ast$}
\end{overpic}
\end{center}
\caption{Smallest enclosing circle for acute triangle $\triangle P^1P^2P^3$}
\label{fig:1}
\end{figure}

\noindent(I) Case that $\triangle P^1P^2P^3$ is acute triangle (see Fig.\ref{fig:1}).

In this case, the circumcenter of $\triangle P^1P^2P^3$, i.e., the equally distant point $Q^0$ from $P^1,P^2,P^3$ is the center of the smallest enclosing circle $\Gamma$ and $d(P^1,Q^0)(=d(P^2,Q^0)=d(P^3,Q^0))$ is its radius $d^\ast$.

From this example, we find it valid to consider the equidistant point $Q^0$ from the given points $P^1,\cdots,P^m$.

\begin{figure}[t]
\begin{center}
\begin{overpic}[width=7.5cm]{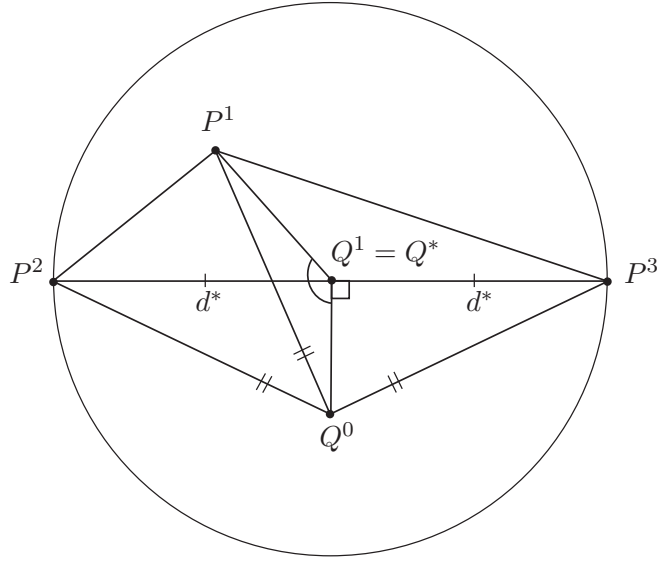}
\put(27,76){$P^1$}
\put(-7,49){$P^2$}
\put(102,49){$P^3$}
\put(48,20){$Q^0$}
\put(50,53){$Q^1=Q^\ast$}
\put(74,44){$d^\ast$}
\put(26,44){$d^\ast$}
\end{overpic}
\end{center}
\caption{Smallest enclosing circle for obtuse triangle $\triangle P^1P^2P^3$}
\label{fig:2}
\end{figure}

\noindent(II) Case that $\triangle P^1P^2P^3$ is obtuse triangle (see Fig.\ref{fig:2}).

In this case, assuming $\angle P^1$ is an obtuse angle, we see that the center $Q^\ast$ of the smallest enclosing circle $\Gamma$ is the midpoint of the side $P^2P^3$, and the radius $d^\ast$ is equal to $d(P^2,Q^\ast)(=d(P^3,Q^\ast))$. 
The equidistant point $Q^0$ from $P^1,P^2,P^3$ exists, however, $Q^0$ is not the center of the smallest enclosing circle because $Q^0$ is outside of $\triangle P^1P^2P^3$. Then, defining $Q^1$ as the nearest point from $Q^0$ on the straight line $L(P^2,P^3)$ passing through $P^2,P^3$, we write it as $Q^1=\pi(Q^0|L(P^2,P^3))$. $Q^1$ is called the projection of $Q^0$ onto $L(P^2,P^3)$. $Q^1$ is the midpoint of the side $P^2P^3$, therefore, we have $Q^\ast=Q^1$. 

From this example, we find it valid to consider the projection of the equidistant point $Q^0$ onto some set if $Q^0$ is outside of $\triangle P^1P^2P^3$.
In this paper, we call it the {\it projection algorithm} that calculates the center $Q^\ast$ of the smallest enclosing circle (or the output distribution $Q^\ast$ achieving the channel capacity) by the projection onto a straight line passing through two points or an affine subspace spanned by plural points.

\subsection{Barycentric coordinate}
Let $O$ be the origin of $\mathbb{R}^n$. Henceforth, for the sake of simplicity, we write $P^i$ instead of $\overrightarrow{OP^i}$, $\lambda_1P^1+\lambda_2P^2$ instead of $\lambda_1\overrightarrow{OP^1}+\lambda_2\overrightarrow{OP^2}$, and $P^2-P^1$ instead of $\overrightarrow{P^1P^2}$, and so on. Depending on the case, we consider $P^i$ as a point in $\mathbb{R}^n$, or as a vector $\overrightarrow{OP^i}$.

We say that $m$ points $P^1,\cdots,P^m\in\mathbb{R}^n$ are {\it in general position} if the vectors $P^2-P^1,\cdots,P^m-P^1$ are linearly independent, or
\begin{align}
\label{eqn:independent}
\text{rank}\left(\begin{array}{c}
P^2-P^1\\
\ \vdots\\
P^m-P^1
\end{array}\right)=m-1.
\end{align}
Let $L_0=L(P^1,\cdots,P^m)$ denote the affine subspace spanned by $P^1,\cdots,P^m$, i.e., the minimum affine subspace including $P^1,\cdots,P^m$, then we have $\dim L_0=m-1$ under the condition (\ref{eqn:independent}). 

We will use the barycentric coordinate to represent the position of a point in $L_0$. Consider $m$ points $P^1,\cdots,P^m\in\mathbb{R}^n$ in general position. The {\it barycentric coordinate} of a point $Q\in L_0$ about $P^1,\cdots,P^m$ is defined as the $m$-tuple of real numbers $\bm{\lambda}=(\lambda_1,\cdots,\lambda_m)$ with 
\begin{numcases}
{}
Q=\lambda_1P^1+\cdots+\lambda_mP^m,\label{eqn:vector}\\
\lambda_1+\cdots+\lambda_m=1.
\end{numcases}
The barycentric coordinate $\bm\lambda$ in the problem of smallest enclosing circle corresponds to the input probability $\bm\lambda$ in the problem of channel capacity.
\subsection{Analysis for smallest enclosing circle}
The smallest enclosing circle $\Gamma(P^1,\cdots,P^m)$ for $P^1,\cdots,$ $P^m\in\mathbb{R}^n$ is obtained by solving the $\min\max$ problem (\ref{eqn:minimaxd}). In (\ref{eqn:minimaxd}), the problem is expressed by the distance $d(P^i,Q)$, so it is easily understood geometrically. Therefore, it is possible to develop a solution algorithm by geometric considerations. In fact, in this paper, we will develop new algorithms based on (\ref{eqn:minimaxd}). However, in order to prove that the algorithm is correct, the double optimization problem as (\ref{eqn:minimaxd}) is difficult to apply. Then, a convex optimization problem (convex programming) as a simple optimization problem equivalent to (\ref{eqn:minimaxd}) is given as follows.

The coordinates of $P^1,\cdots,P^m$ are defined by
\begin{align}
\label{eqn:Picoordinate}
P^i=(P^i_1,\cdots,P^i_n)\in\mathbb{R}^n,\,i=1,\cdots,m,
\end{align}
and a matrix $\Phi$ with row vectors $P^i$ is defined by
\begin{align}
\label{eqn:matrixPhi}
\Phi=\left(\begin{array}{c}P^1\\\vdots\\P^m\end{array}\right)
=\left(\begin{array}{ccc}P^1_1 & \cdots & P^1_n\\
\vdots && \vdots\\
P^m_1 & \cdots & P^m_n\end{array}\right)\in\mathbb{R}^{m\times n}.
\end{align}
A vector $\bm{a}$ is defined by $\bm{a}=(\|P^1\|^2,\cdots,\|P^m\|^2)$, where $\|P^i\|^2=\sum_{j=1}^n(P^i_j)^2$ is the squared norm of the vector $P^i,\,i=1,\cdots,m$. Then, a function $f(\bm\lambda,\Phi)$ of $\bm{\lambda}\in\mathbb{R}^m$ associated with $\Phi$ is defined by 
\begin{align}
f(\bm{\lambda},\Phi)=\bm{\lambda}\,{^t}\!\bm{a}-\bm{\lambda}\Phi\,{^t}\Phi\,{^t}\hspace{-0.3mm}\bm{\lambda},
\end{align}
where, $^t$ denotes the transposition of vector or matrix. $f(\bm\lambda,\Phi)$ is a differentiable and convex upward function of $\bm\lambda$. Let us define $\bar\Delta^m$ by
\begin{align*}
\bar\Delta^m=\{\bm\lambda=(\lambda_1,\cdots,\lambda_m)|\lambda_i\geq0,i=1,\cdots,m,\sum_{i=1}^m\lambda_i=1\},
\end{align*}
then the convex optimization problem
\begin{align}
\label{eqn:cp}
\max_{\bm{\lambda}\in\bar{\Delta}^m}f(\bm{\lambda},\Phi)
\end{align}
is equivalent to the problem of smallest enclosing circle \cite{sch}. For $\bm{\lambda}=\bm{\lambda}^\ast$ that achieves (\ref{eqn:cp}), $Q^\ast\equiv\bm\lambda^\ast\Phi$ is the center of the smallest enclosing circle $\Gamma(P^1,\cdots,P^m)$ and $d^\ast\equiv\sqrt{f(\bm{\lambda}^\ast,\Phi)}$ is its radius \cite{sch}. 

For (\ref{eqn:cp}), the following Kuhn-Tucker condition holds \cite{gal},\cite{sch}. 

{\bf Theorem:} (Kuhn-Tucker condition for the problem of smallest enclosing circle) A necessary and sufficient condition for $\bm\lambda^\ast=(\lambda^\ast_1,\cdots,\lambda^\ast_m)\in\bar{\Delta}^m$ to achieve (\ref{eqn:cp}) is that there exists $d_0$ with
\begin{align}
\label{eqn:KT}
d(P^i,\bm\lambda^\ast\Phi)\left\{\begin{array}{ll}=d_0, & {\mbox{\rm for}}\ i\ {\mbox{\rm with}}\ \lambda^\ast_i>0,\\
\leq d_0, & {\mbox{\rm for}}\ i\ {\mbox{\rm with}}\ \lambda^\ast_i=0.
\end{array}\right.
\end{align}
Then, $Q^\ast=\bm\lambda^\ast\Phi$ is the center of the smallest enclosing circle $\Gamma(P^1,\cdots,P^m)$ and $d^\ast=d_0$ is its radius.
\subsection{Equidistant point from $P^1,\cdots,P^m$ and its barycentric coordinate}
\label{sec:quidistantpointanditsbarycentriccoordinate}
For $P^1,\cdots,P^m\in\mathbb{R}^n$, the {\it equidistant point} from $P^1,\cdots,$ $P^m$ is a point $Q^0\in L_0=L(P^1,\cdots,P^m)$ that satisfies $d(P^1,Q^0)=\cdots=d(P^m,Q^0)$. 

Now, we assume in this chapter that $P^1\cdots,P^m$ are in general position. Then, defining 
\begin{align}
\Psi=\left(\begin{array}{c}
P^2-P^1\\
\ \vdots\\
P^m-P^1
\end{array}\right)\in\mathbb{R}^{(m-1)\times n},
\end{align}
we have from (\ref{eqn:independent})
\begin{align}
\label{eqn:rankPsim-1}
\text{rank}\,\Psi=m-1.
\end{align}
We will calculate the barycentric coordinate $\bm\lambda^0$ of the equidistant point $Q^0\in L_0$.

\subsubsection{Calculation of $\bm\lambda^0$}
The coordinate of $P^i$ is defined by (\ref{eqn:Picoordinate}) and the $m\times n$ matrix $\Phi$ is defined by (\ref{eqn:matrixPhi}). Define the coordinate of $Q^0$ by
\begin{align}
Q^0&=(Q^0_1,\cdots,Q^0_n)\in\mathbb{R}^n,
\end{align}
and further define the following:
\begin{align}
\hat{P}^i&=(1,P^i)\nonumber\\
&=(1,P^i_1,\cdots,P^i_n)\in\mathbb{R}^{n+1},\,i=1,\cdots,m,\label{eqn:hatP}
\end{align}
\begin{align}
\hat{Q}^0&=(1,Q^0)=(1,Q^0_1,\cdots,Q^0_n)\in\mathbb{R}^{n+1},\label{eqn:hatQ}
\end{align}
\begin{align}
\hat\Phi&=\left(\begin{array}{c}
\hat{P}^1\\
\ \vdots\\
\hat{P}^m
\end{array}\right)\nonumber\\
&=\left(\begin{array}{cccc}
1 & P^1_1 & \cdots & P^1_n\\
\vdots & \vdots & & \vdots\\
1 & P^m_1 & \cdots & P^m_n
\end{array}\right)\in\mathbb{R}^{m\times(n+1)},
\end{align}
\begin{align}
J&=\left(\begin{array}{ccccc}
-1 & 1 & 0 & \cdots & 0\\
-1 & 0 & 1 & \cdots & 0\\
\vdots & \vdots & \vdots & \ddots & \vdots\\
-1 & 0 & 0 & \cdots & 1\\
\end{array}\right)\in\mathbb{R}^{(m-1)\times m},
\end{align}
\begin{align}
\hat{\bm{a}}&=(\|\hat{P}^1\|^2,\cdots,\|\hat{P}^m\|^2)\in\mathbb{R}^m,
\end{align}
\begin{align}
\bm{1}&=(1,\cdots,1)\in\mathbb{R}^m.
\end{align}
Because $P^1,\cdots,P^m\in\mathbb{R}^n$ are in general position, we see $\mbox{\rm rank}\,\hat\Phi=m$. In fact, if $\sum_{i=1}^mc_i\hat{P}^i=O$, then $\sum_{i=2}^mc_i(P^i-P^1)=O$, hence from (\ref{eqn:rankPsim-1}) we have $c_i=0,\,i=1,\cdots,m$.

The barycentric coordinate $\bm\lambda^0$ of $Q^0$ about $P^1,\cdots,P^m$ satisfies
\begin{numcases}
{}
Q^0=\bm\lambda^0\Phi,\label{eqn:barycentric7}\\
\bm\lambda^0\hspace{0.3mm}{^t}\hspace{-0.3mm}\bm{1}=1.\label{eqn:barycentric8}
\end{numcases}
Together (\ref{eqn:barycentric7}) and (\ref{eqn:barycentric8}) is written as
\begin{align}
\hat{Q}^0=\bm\lambda^0\hat\Phi.\label{eqn:barycentric9}
\end{align}
The equidistant point $Q^0$ from $P^1,\cdots,P^m$ satisfies
\begin{align}
d(P^i,Q^0)=d(P^1,Q^0),\,i=2,\cdots,m,\label{eqn:toukyori5}
\end{align}
hence we have from (\ref{eqn:toukyori5})
\begin{align}
2(P^i-P^1)\hspace{0.3mm}{^t}\hspace{-0.3mm}Q^0=\|P^i\|^2-\|P^1\|^2,\,i=2,\cdots,m,\label{eqn:toukyori6}
\end{align}
and from (\ref{eqn:hatP}), (\ref{eqn:hatQ}), (\ref{eqn:toukyori6}),
\begin{align}
2(\hat{P}^i-\hat{P}^1)\hspace{0.3mm}{^t}\hspace{-0.3mm}\hat{Q}^0=\|\hat{P}^i\|^2-\|\hat{P}^1\|^2,\,i=2,\cdots,m.\label{eqn:toukyori7}
\end{align}
Since (\ref{eqn:toukyori7}) is written as 
\begin{align}
2J\hat\Phi\hspace{0.7mm}{^t}\hspace{-0.3mm}\hat{Q}^0=J\hspace{0.3mm}{^t}\hspace{-0.3mm}\hat{\bm{a}},\label{eqn:toukyori8}
\end{align}
we have from (\ref{eqn:barycentric9})
\begin{align}
2J\hat\Phi\hspace{0.8mm}{^t}\hspace{-0.3mm}\hat\Phi\hspace{0.3mm}{^t}\hspace{-0.3mm}\bm\lambda^0
=J\hspace{0.3mm}{^t}\hspace{-0.3mm}\hat{\bm{a}}.\label{eqn:toukyori9}
\end{align}
Defining $M\equiv\hat\Phi\hspace{0.7mm}{^t}\hspace{-0.3mm}\hat\Phi\in\mathbb{R}^{m\times m}$, we have from (\ref{eqn:toukyori9})
\begin{align}
J(2M\hspace{0.3mm}{^t}\hspace{-0.3mm}\bm\lambda^0-\hspace{0.3mm}{^t}\hspace{-0.3mm}\hat{\bm{a}})
=\bm{0}.\label{eqn:toukyori10}
\end{align}
Because $\text{rank}\,M=\text{rank}\,\hat\Phi=m$, $M$ is non-singular. Since Ker\,$J=\{\tau\hspace{0.3mm}{^t}\hspace{-0.3mm}\bm{1}|\tau\in\mathbb{R}\}$, from (\ref{eqn:toukyori10}) there exists $\tau\in\mathbb{R}$ with
\begin{align}
\bm\lambda^0=\ds\frac{1}{2}(\hat{\bm{a}}+\tau\bm{1})M^{-1}.\label{eqn:barycentric10}
\end{align}
So, from (\ref{eqn:barycentric8})
\begin{align}
1=\bm\lambda^0\hspace{0.3mm}{^t}\hspace{-0.3mm}\bm{1}
=\ds\frac{1}{2}(\hat{\bm{a}}+\tau\bm{1})M^{-1}\hspace{0.3mm}{^t}\hspace{-0.3mm}\bm{1},\label{eqn:barycentric11}
\end{align}
and from (\ref{eqn:barycentric11}), we have
\begin{align}
\tau=\ds\frac{2-\hat{\bm{a}}M^{-1}\hspace{0.3mm}{^t}\hspace{-0.3mm}\bm{1}}{\bm{1}M^{-1}\hspace{0.3mm}{^t}\hspace{-0.3mm}\bm{1}}.\label{eqn:tau1}
\end{align}
Substituting (\ref{eqn:tau1}) into (\ref{eqn:barycentric10}), we finally have
\begin{align}
\label{eqn:barycentric12}
\bm\lambda^0=\ds\frac{1}{2}\left(\hat{\bm{a}}+
\ds\frac{2-\hat{\bm{a}}M^{-1}\hspace{0.3mm}{^t}\hspace{-0.3mm}\bm{1}}{\bm{1}M^{-1}\hspace{0.3mm}{^t}\hspace{-0.3mm}\bm{1}}\bm{1}\right)M^{-1}.
\end{align}
This is the barycentric coordinate of the equidistant point $Q^0$.

\subsection{Inner product, Pythagorean theorem and projection in $\mathbb{R}^n$}
We will describe the inner product, Pythagorean theorem and projection in $\mathbb{R}^n$, which are important to determine the solution of the problem of smallest enclosing circle by the projection algorithm.
\subsubsection{Inner product}
For three points $Q^k=(Q^k_1,\cdots,Q^k_n)\in\mathbb{R}^n,\,k=1,2,3$, let us define the inner product $(Q^1-Q^2,Q^3-Q^2)$ by
\begin{align}
(Q^1-Q^2,Q^3-Q^2)=\ds\sum_{j=1}^n(Q^1_j-Q^2_j)(Q^3_j-Q^2_j).
\end{align}
This is the inner product of two vectors $Q^1-Q^2(=\overrightarrow{Q^2Q^1})$ and $Q^3-Q^2(=\overrightarrow{Q^2Q^3})$.

We have the following lemmas.
\begin{lemma}
\label{lem:1}
For points $P^i\,(i=1,\cdots,m),Q,R$ in $\mathbb{R}^n$, consider the inner products $\sigma_i=(P^i-Q,R-Q),\,i=1,\cdots,m$. If $\sum_{i=1}^m\lambda_i=1$, then we have
\begin{align*}
\sum_{i=1}^m\lambda_i\sigma_i=\Big(\sum_{i=1}^m\lambda_iP^i-Q,R-Q\Big).
\end{align*}
\end{lemma}
{\bf Proof}: By a simple calculation.\hfill$\Box$

\begin{lemma}
\label{lem:2}
For any $P,\,Q\in\mathbb{R}^n,\,P\neq Q$, we have
\begin{align*}
(P-Q,P-Q)>0.
\end{align*}
\end{lemma}
{\bf Proof}: By a simple calculation.\hfill$\Box$
\begin{lemma}
\label{lem:3}
For any $P,Q,R\in\mathbb{R}^n$, we have
\begin{align*}
2(P-Q,R-Q)=d^2(P,Q)+d^2(Q,R)-d^2(P,R).
\end{align*}
\end{lemma}
{\bf Proof}: By a simple calculation.\hfill$\Box$
\subsubsection{Pythagorean theorem}
For three points $P,Q,R$ in $\mathbb{R}^n$, the following Pythagorean theorem and its inequality versions hold.
\begin{theorem}$\rm(Pythagorean)$ 
\label{theorem:1}
For $P,Q,R\in\mathbb{R}^n$, we have
%
%
%
\begin{align}
(P-Q,R-Q)\gtreqqless0 \iff d^2(P,Q)+d^2(Q,R)\gtreqqless d^2(P,R).
\end{align}
\end{theorem}
{\bf Proof}: By Lemma \ref{lem:3}.\hfill$\Box$
\subsubsection{Projection}
For a point $Q'\in\mathbb{R}^n$ and a subset $L\subset\mathbb{R}^n$, the point $Q=Q''$ that achieves $\min_{Q\in L}d(Q,Q')$ is called the {\it projection} of $Q'$ onto $L$, and is denoted by by $Q''=\pi(Q'|L)$. In this paper, we consider only affine subspaces of $\mathbb{R}^n$ as $L$.
\begin{lemma}
\label{lem:4}
Let $L$ be an affine subspace of $\mathbb{R}^n$. For any $Q'\in\mathbb{R}^n$, the projection $Q''=\pi(Q'|L)$ exists and is unique. Moreover, $Q''=\pi(Q'|L)$ is equivalent to that $(P-Q'',Q'-Q'')=0$ holds for any $P\in L$.
\end{lemma}
{\bf Proof}: see \cite{str}.\hfill$\Box$
\subsubsection{Projection of equidistant point}
For $P^1,\cdots,P^m\in\mathbb{R}^n$ in general position, let $L_0=L(P^1,\cdots,P^m)$ be the affine subspace spanned by $P^1,\cdots,P^m$ and $Q^0\in L_0$ be the equidistant point from $P^1,\cdots,P^m$. We see from (\ref{eqn:barycentric12}) that $Q^0$ uniquely exists. Further, we define $L_k=L(P^{k+1},\cdots,P^m),\,k=0,\cdots,m-2$ as the affine subspace spanned by $P^{k+1},\cdots,P^m$. $L_0\supset L_1\supset\cdots\supset L_k\supset\cdots$ is a decreasing sequence of affine subspaces whose dimensions are decreasing by $1$.

Let $Q^1=\pi(Q^0|L_1)$ be the projection of $Q^0$ onto $L_1$. Further, let $Q^k=\pi(Q^{k-1}|L_k),\,k=1,\cdots,m-2$. We have the following lemmas.
\begin{lemma}
\label{lem:5}
$Q^k=\pi(Q^0|L_k),\,k=0,1,\cdots,m-2.$
\end{lemma}
{\bf Proof}: 
It is trivial for $k=0,1$ by definition. Then, we prove for $k=2$. From Lemma \ref{lem:4} and Theorem \ref{theorem:1}, we have $d^2(Q,Q^0)=d^2(Q,Q^1)+d^2(Q^1,Q^0)$ for any $Q\in L_2\subset L_1$. Therefore, with respect to $Q\in L_2$ minimizing $d(Q,Q^0)$ is equivalent to minimizing $d(Q,Q^1)$. Because the projections $\pi(Q^0|L_2)$ and $\pi(Q^1|L_2)$ are unique by Lemma \ref{lem:4}, so we have $\pi(Q^0|L_2)=\pi(Q^1|L_2)$. For $k\geq3$, we can prove it by mathematical induction.\hfill$\Box$

\begin{lemma}
\label{lem:6}
$(Q^i-Q^k,Q^0-Q^k)=(Q^i-Q^k,Q^i-Q^k),\,i=0,1,\cdots,k,k=0,1,\cdots,m-2.$
\end{lemma}
{\bf Proof}: 
By calculation, we have $(Q^i-Q^k,Q^0-Q^k)=-(Q^k-Q^i,Q^0-Q^i)+(Q^i-Q^k,Q^i-Q^k)$. Since $Q^k\in L_k\subset L_i$, we have $(Q^k-Q^i,Q^0-Q^i)=0$ by Lemma \ref{lem:4}.\hfill$\Box$

\subsection{Search for $Q^\ast$ by projection algorithm}
\label{subsec:tansaku1}
As we discussed in section \ref{subsec:eikakudonkaku}, it is valid to consider the equidistant point $Q^0$ and the projection of $Q^0$ onto some affine subspace. We are calling this a projection algorithm.

For given points $P^1,\cdots,P^m$, let $Q^\ast$ be the center of the smallest enclosing circle $\Gamma=\Gamma(P^1,\cdots,P^m)$ and $d^\ast$ be its radius. In this chapter, we are assuming that $P^1,\cdots,P^m$ are in general position, and then we will show that $Q^\ast,\,d^\ast$ are calculated by the projection algorithm in some situations.
\subsection{Situation 1 $[$There is just one negative component of barycentric coordinate at every projection.$]$}
\label{subsec:joukyou1}
First, let us consider the following example.
\begin{example}
\label{example:1}
{\rm 
Consider four points $P^1=(-10, 1, -3), P^2=(-9, -2, 8), P^3=(-8, 10, -5), P^4=(4, -8, 8)$ given in $\mathbb{R}^3$. $P^1,\cdots,P^4$ are in general position. The equidistant point from $P^1,\cdots,P^4$ is $Q^0=(1.93,4.59,2.85)$, and its barycentric coordinate $\bm\lambda^0$ is
\begin{align}
\label{eqn:joukyou1-1}
\bm\lambda^0=(-0.84,0.04,1.11,0.69),
\end{align}
which is calculated by (\ref{eqn:barycentric12}). Since $\lambda^0_1=-0.84<0$, we remove $P^1$ and calculate the projection of $Q^0$ onto $L_1=L(P^2,P^3,P^4)$, i.e., $Q^1=\pi(Q^0|L_1)$. We have $Q^1=(0.48,1.45,-0.16)$ and its barycentric coordinate $\bm\lambda^1$ is 
\begin{align}
\label{eqn:joukyou1-2}
\bm\lambda^1=(0,-0.31,0.63,0.68).
\end{align}
Since $\lambda^1_2=-0.31<0$, we remove $P^2$ and calculate the projection $Q^2$ of $Q^1$ onto $L_2=L(P^3,P^4)$, i.e., $Q^2=\pi(Q^1|L_2)$. We have $Q^2=(-2,1,1.5)$, and its barycentric coordinate $\bm\lambda^2$ is 
\begin{align}
\label{eqn:joukyou1-3}
\bm\lambda^2=(0,0,0.5,0.5).
\end{align}
Then, we have $d(P^1,Q^2)=9.18$, $d(P^2,Q^2)=10.01$, $d(P^3,Q^2)=d(P^4,Q^2)=12.62$, so by the Kuhn-Tucker condition (\ref{eqn:KT}), we see that $Q^2$ is the center of the smallest enclosing circle, i.e., $Q^\ast=Q^2$, and thus, $d^\ast=d(P^3,Q^2)$.
}
\end{example}

\medskip

In this section, we represent ``situation 1'' as the case that there is just one negative component of barycentric coordinate at every projection like (\ref{eqn:joukyou1-1}),\,(\ref{eqn:joukyou1-2}), and all the components are non-negative at the last projection like (\ref{eqn:joukyou1-3}). We will calculate below the smallest enclosing circle by the projection algorithm in situation 1. 
\medskip

\noindent\underline{Assumption of situation 1}\ \ 
Assume $P^1,\cdots,P^m\in\mathbb{R}^n$ are in general position, and let $L_k=L(P^{k+1},\cdots,P^m),\,k=0,1,\cdots,$ $m-2$ be the affine subspace spanned by $P^{k+1},$ $\cdots,P^m$. Let $Q^0\in L_0$ be the equidistant point from $P^1,\cdots,P^m$, and define $Q^k=\pi(Q^{k-1}|L_k),\,k=1,\cdots,m-2$. The barycentric coordinate of $Q^k$ about $P^1,\cdots,P^m$ is denoted by $\bm\lambda^k=(\lambda^k_1,\cdots,\lambda^k_m)$. Let $K=0,1,\cdots,m-2$. We assume that for $k=0,1,\cdots,K-1$, there is just one negative component of $\bm\lambda^k$, and for $k=K$, all the components of $\bm\lambda^K$ are non-negative. That is, 
%
%
%
\begin{align*}
&\bm{\lambda}^0=(\lambda^0_1,\lambda^0_2,\cdots,\lambda^0_m)=(-,+,\cdots,+),\\
&\bm\lambda^1=(\lambda^1_1,\lambda^1_2,\lambda^1_3,\cdots,\lambda^1_m)=(0,-,+,\cdots,+),\\
&\bm\lambda^2=(\lambda^2_1,\lambda^2_2,\lambda^2_3,\lambda^2_4,\cdots,\lambda^2_m)=(0,0,-,+,\cdots,+),\\
&\hspace{5mm}\vdots\\
&\bm\lambda^{K-1}=(\lambda^{K-1}_1,\cdots,\lambda^{K-1}_{K-1},\lambda^{K-1}_K,\cdots,\lambda^{K-1}_m)=(0,\cdots,0,-,+,\cdots,+),\\
&\bm\lambda^K=(\lambda^K_1,\cdots,\lambda^K_K,\lambda^K_{K+1},\cdots,\lambda^K_m)=(0,\cdots,0,+,\cdots,+).
\end{align*}
\noindent If $K=0$, we assume $\bm\lambda^0=(+,+,\cdots,+)$. Summarizing above, we have, 
for $k=0,1,\cdots,K-1$,
\begin{subnumcases}
{\lambda^k_i}
=0, & \hspace{-4mm}$i=1,\cdots,k,$\label{eqn:1a}\\[-1mm]
<0, & \hspace{-4mm}$i=k+1,$\label{eqn:1b}\\[-1mm]
>0, & \hspace{-4mm}$i=k+2,\cdots,m,$\label{eqn:1c}
\end{subnumcases}
and for $k=K$,
\begin{subnumcases}
{\lambda^K_i}
=0, & \hspace{-4mm}$i=1,\cdots,K,$\label{eqn:2a}\\[-1mm]
>0, & \hspace{-4mm}$i=K+1,\cdots,m$.\label{eqn:2b}
\end{subnumcases}
\noindent(The end of Assumption of situation 1)

\medskip

Now, let us consider the inner products $\sigma_i\equiv(P^i-Q^K,Q^0-Q^K),\,i=1,\cdots,m$. We have the following lemma.

\begin{lemma}
\label{lem:7}
$\sigma_i\left\{\begin{array}{ll}
<0, & i=1,\cdots,K,\\
=0, & i=K+1,\cdots,m.
\end{array}\right.$
\end{lemma}
{\bf Proof}: 
By Lemma \ref{lem:5}, we have $Q^K=\pi(Q^0|L_K),\,L_K=L(P^{K+1},\cdots,P^m)$. Since $P^i\in L_K$ for $i=K+1,\cdots,m$, so by Lemma \ref{lem:4} we have $P^i-Q^K\perp Q^0-Q^K$, thus
\begin{align}
\sigma_i=0,\,i=K+1,\cdots,m.\label{eqn:sigma_i2}
\end{align}
Next, we will prove $\sigma_i<0,\,i=1,\cdots,K$ by mathematical induction in the order of $i=K,K-1,\cdots,1$.

\noindent(I) Prove $\sigma_K<0$:

From (\ref{eqn:1a}), (\ref{eqn:sigma_i2}), we have
\begin{align}
\lambda^{K-1}_K\sigma_K&=\sum_{i=1}^m\lambda^{K-1}_i\sigma_i\nonumber\\
&=\left(\sum_{i=1}^m\lambda^{K-1}_iP^i-Q^K,Q^0-Q^K\right)\ \ \mbox{\rm(by Lemma \ref{lem:1})}\nonumber\\
&=(Q^{K-1}-Q^K,Q^0-Q^K)\nonumber\\
&=(Q^{K-1}-Q^K,Q^{K-1}-Q^K)\ \ \mbox{\rm(by Lemma \ref{lem:6})}\nonumber\\
&>0.\ \ \mbox{\rm(by Lemma \ref{lem:2})}\label{eqn:6}
\end{align}
Since $\lambda^{K-1}_K<0$ from (\ref{eqn:1b}), we have $\sigma_K<0$.

\noindent(II) Assuming $\sigma_K<0,\,\sigma_{K-1}<0,\cdots,\sigma_{k+1}<0$, prove $\sigma_k<0$:

From (\ref{eqn:1a}), (\ref{eqn:sigma_i2}), we have
\begin{align}
\sum_{i=k}^K\lambda^{k-1}_i\sigma_i&=\sum_{i=1}^m\lambda^{k-1}_i\sigma_i\nonumber\\
&=\left(\sum_{i=1}^m\lambda^{k-1}_iP^i-Q^K,Q^0-Q^K\right)\nonumber\\
&=(Q^{k-1}-Q^K,Q^0-Q^K)\nonumber\\
&=(Q^{k-1}-Q^K,Q^{k-1}-Q^K)\nonumber\\
&>0,\label{eqn:9}
\end{align}
i.e., $\lambda^{k-1}_k\sigma_k+\sum_{i=k+1}^K\lambda^{k-1}_i\sigma_i>0$. By the induction hypothesis, $\sigma_i<0,\,i=k+1,\cdots,K$, and by (\ref{eqn:1b}), $\lambda^{k-1}_k<0$, by (\ref{eqn:1c}), $\lambda^{k-1}_i>0,\,i=k+1,\cdots,K$, therefore we obtain $\sigma_k<0$.

\noindent(III) From the above (I),(II), we have $\sigma_i<0,\,i=1,\cdots,K$.\hfill$\Box$

\begin{lemma}
\label{lem:8}
$d(P^i,Q^K)<d(P^{K+1},Q^K)=\cdots=d(P^m,$ $Q^K),\,i=1,\cdots,K$.
\end{lemma}
{\bf Proof}: 
By Lemma \ref{lem:7}, $\sigma_i=0,\,i=K+1,\cdots,m$, thus by Theorem \ref{theorem:1}, we have $d(P^{K+1},Q^K)=\cdots=d(P^m,Q^K)$. Also by Lemma \ref{lem:7}, $\sigma_i<0,\,i=1,\cdots,K$, thus for $i=1,\cdots,K$, we have by Theorem \ref{theorem:1},
\begin{align*}
d^2(P^i,Q^K)&<d^2(P^i,Q^0)-d^2(Q^K,Q^0)\\
&=d^2(P^{K+1},Q^0)-d^2(Q^K,Q^0)\\
&=d^2(P^{K+1},Q^K).
\end{align*}\hfill$\Box$
\begin{theorem}
\label{theorem:2}
The center of the smallest enclosing circle $\Gamma$ is $Q^\ast=Q^K$, and its radius is $d^\ast=d(P^{K+1},Q^K)$.
\end{theorem}
{\bf Proof}: 
It follows from (\ref{eqn:2a}), (\ref{eqn:2b}), Lemma \ref{lem:8} and the Kuhn-Tucker condition (\ref{eqn:KT}).\hfill$\Box$

\subsection{Situation 2 $[$$m=2,3,4$ and $n$ is arbitrary.$]$}
We will calculate by the projection algorithm the center $Q^\ast$ and the radius $d^\ast$ of the smallest enclosing circle $\Gamma=\Gamma(P^1,\cdots,P^m)$ in the case that $m=2,3,4$ and $n$ is arbitrary. Of course our goal is to find an algorithm to calculate $Q^\ast$ and $d^\ast$ for every $m$, but at present it is possible to solve only for $m=2,3,4$.
\subsubsection{Case of $m=2$}
Consider the smallest enclosing circle $\Gamma=\Gamma(P^1,P^2)$ for two different points $P^1,P^2$ in ${\mathbb R}^n$. Let $Q^0$ be the midpoint of the line segment $P^1P^2$, then we have
\begin{theorem}
\label{theorem:3}
The center of \,$\Gamma$ is $Q^\ast=Q^0$ and the radius is $d^\ast=d(P^1,Q^0)$.
\end{theorem}
\subsubsection{Case of $m=3$}
\label{subsubsec:m=3}
Consider the smallest enclosing circle $\Gamma=\Gamma(P^1,P^2,P^3)$ for three points $P^1,P^2,P^3$ in general position in ${\mathbb R}^n$. Let $L_0=L(P^1,P^2,P^3)$ be the affine subspace spanned by $P^1,P^2,P^3$, and $Q^0\in L_0$ be the equidistant point from $P^1,P^2,P^3$. The barycentric coordinate of $Q^0$ about $P^1,P^2,P^3$ is denoted by ${\bm\lambda}^0=(\lambda^0_1,\lambda^0_2,\lambda^0_3)$.

Based on the signs of the components of $\bm\lambda^0$, we can make the following classification without loss of generality.

\medskip

Case 3-1: $\lambda^0_1\geq0,\lambda^0_2\geq0,\lambda^0_3\geq0$

\medskip

Case 3-2: $\lambda^0_1<0,\lambda^0_2\geq0,\lambda^0_3\geq0$

\medskip

Since $Q^0$ is the equidistant point from $P^1,P^2,P^3$, there must be at least two positive components in the barycentric coordinate of $Q^0$. Therefore, all the cases are exhausted by Case 3-1 and Case 3-2. For each case, we will determine the center $Q^\ast$ and the radius $d^\ast$ of the smallest enclosing circle $\Gamma=\Gamma(P^1,P^2,P^3)$. 

\medskip
\noindent{\bf Case 3-1}\\
In this case, $\triangle P^1P^2P^3$ is an acute triangle. We have the following theorem by the Kuhn-Tucker condition (\ref{eqn:KT}).
\begin{theorem}
\label{theorem:4}
The center of \,$\Gamma$ is $Q^\ast=Q^0$ and the radius is $d^\ast=d(P^1,Q^0)$.
\end{theorem}
\medskip
\noindent{\bf Case 3-2}\\
In this case, $\triangle P^1P^2P^3$ is an obtuse triangle. Let $L_1=L(P^2,P^3)$ be the line connecting the two points $P^2$, $P^3$. By the assumption $\lambda^0_1<0$, the equidistant point $Q^0$ is in the opposite side of $P^1$ with respect to $L_1$. Let $Q^1=\pi(Q^0|L_1)$ be the projection of $Q^0$ onto $L_1$, then $Q^1$ is the midpoint of $P^2P^3$. Denoting by $\bm\lambda^1=(\lambda^1_1,\lambda^1_2,\lambda^1_3)$ the barycentric coordinate of $Q^1$ about $P^1, P^2, P^3$, we have $\lambda^1_1=0, \lambda^1_2=\lambda^1_3=1/2$, hence this case is the situation 1. So, we have
\begin{theorem}
\label{theorem:5}
The center of \,$\Gamma$ is $Q^\ast=Q^1$ and the radius is $d^\ast=d(P^2,Q^1)$.
\end{theorem}
\subsubsection{Case of $m=4$}
\label{subsec:circlem4}
Consider the smallest enclosing circle $\Gamma=\Gamma(P^1,P^2,P^3,P^4)$ for four points $P^1,P^2,P^3,P^4$ in general position in ${\mathbb R}^n$. Let $L_0=L(P^1,\cdots,P^4)$ be the affine subspace spanned by $P^1,\cdots,P^4$ and $Q^0\in L_0$ be the equidistant point from $P^1,\cdots,P^4$. The barycentric coordinate of $Q^0$ about $P^1,\cdots,P^4$ is denoted by $\bm\lambda^0=(\lambda^0_1,\cdots,\lambda^0_4)$. Without loss of generality, we have the following exhaustive classification:

\medskip

Case 4-1: $\lambda^0_1\geq0,\lambda^0_2\geq0, \lambda^0_3\geq0, \lambda^0_4\geq0$

\medskip

Case 4-2: $\lambda^0_1<0,\lambda^0_2\geq0, \lambda^0_3\geq0, \lambda^0_4\geq0$

\medskip

Case 4-3: $\lambda^0_1<0,\lambda^0_2<0, \lambda^0_3\geq0, \lambda^0_4\geq0$

\medskip
\medskip
\noindent{\bf Case 4-1}\\
By the Kuhn-Tucker condition (\ref{eqn:KT}), we have
\begin{theorem}
\label{theorem:6}
The center of \,$\Gamma$ is $Q^\ast=Q^0$ and the radius is $d^\ast=d(P^1,Q^0)$.
\end{theorem}
\medskip
\noindent{\bf Case 4-2}\\
Because $\lambda^0_1<0$, we consider the projection $Q^1=\pi(Q^0|L_1)$ of $Q^0$ onto $L_1=L(P^2,P^3,P^4)$. Let $\bm\lambda^1=(\lambda^1_1,\cdots,\lambda^1_4)$ be the barycentric coordinate of $Q^1$ about $P^1,\cdots,P^4$. Since $P^1\notin L_1$, we have $\lambda^1_1=0$. Based on the signs of the components of $\bm\lambda^1$, we have the following classification without loss of generality.

\medskip

Case 4-2-1: $\lambda^1_1=0,\lambda^1_2\geq0, \lambda^1_3\geq0, \lambda^1_4\geq0$

\medskip

Case 4-2-2: $\lambda^1_1=0,\lambda^1_2<0, \lambda^1_3\geq0, \lambda^1_4\geq0$

\medskip

\medskip
\noindent{\bf Case 4-2-1}\\
This case is the situation 1, so we have 
\begin{theorem}
\label{theorem:7}
The center of \,$\Gamma$ is $Q^\ast=Q^1$ and the radius is $d^\ast=d(P^2,Q^1)$.
\end{theorem}

\medskip
\noindent{\bf Case 4-2-2}\\
Let $Q^2=\pi(Q^1|L_2)$ with $L_2=L(P^3,P^4)$, then $Q^2$ is the midpoint of $P^3P^4$. Denoting by $\bm\lambda^2=(\lambda^2_1,\cdots,\lambda^2_4)$ the barycentric coordinate of $Q^2$ about $P^1,\cdots,P^4$, we have $\lambda^2_1=\lambda^2_2=0,\lambda^2_3=\lambda^2_4=1/2$, hence also this case is the situation 1. So, we have
\begin{theorem}
\label{theorem:8}
The center of \,$\Gamma$ is $Q^\ast=Q^2$ and 
the radius is $d^\ast=d(P^3,Q^2)$.
\end{theorem}

\medskip

This completes Case 4-2, then let us consider Case 4-3.

\medskip
\noindent{\bf Case 4-3}\\
For the equidistant point $Q^0\in L_0$, let us define $Q^{1(1)}=\pi(Q^0|L(P^2,P^3,P^4))$ and $Q^{1(2)}=\pi(Q^0|L(P^1,P^3,P^4))$. Then, denote by $\bm\lambda^{1(1)}=(\lambda^{1(1)}_1,\cdots,\lambda^{1(1)}_4)$ and $\bm\lambda^{1(2)}=(\lambda^{1(2)}_1,\cdots,\lambda^{1(2)}_4)$ the barycentric coordinates of $Q^{1(1)}$ and $Q^{1(2)}$ about $P^1,\cdots,P^4$, respectively. Because $P^1\notin L(P^2,P^3,P^4)$, $P^2\notin L(P^1,P^3,P^4)$, we have
\begin{align}
\lambda^{1(1)}_1=0,\ \lambda^{1(2)}_2=0.\label{eqn:null1}
\end{align}
Then, consider the inner products
\begin{align}
&\sigma^{1(1)}_i=(P^i-Q^{1(1)},Q^0-Q^{1(1)}),\ i=1,\cdots,4,\label{eqn:inner3}\\
&\sigma^{1(2)}_i=(P^i-Q^{1(2)},Q^0-Q^{1(2)}),\ i=1,\cdots,4.\label{eqn:inner4}
\end{align}
\begin{lemma}
\label{lem:13}

\ \ $\sigma^{1(1)}_i\left\{\begin{array}{ll}
<0, & i=1,\\
=0, & i=2,3,4,
\end{array}\right.$ 
$\sigma^{1(2)}_i\left\{\begin{array}{ll}
<0, & i=2,\\
=0, & i=1,3,4.
\end{array}\right.$
\end{lemma}
\medskip
{\bf Proof}: 
From $P^i-Q^{1(1)}\perp Q^0-Q^{1(1)},\,i=2,3,4$, we have $\sigma^{1(1)}_i=0,\,i=2,3,4$. Thus, 
\begin{align}
\lambda^0_1\sigma^{1(1)}_1&=\sum_{i=1}^4\lambda^0_i\sigma^{1(1)}_i\nonumber\\
&=(Q^0-Q^{1(1)},Q^0-Q^{1(1)})>0\label{eqn:lamal2}
\end{align}
holds, and by the assumption $\lambda^0_1<0$ we obtain $\sigma^{1(1)}_1<0$. We can prove for $\sigma^{1(2)}_i$ similarly.\hfill$\Box$

\begin{lemma}
\label{lem:14}
$\lambda^{1(2)}_1\sigma^{1(1)}_1+\lambda^{1(1)}_2\sigma^{1(2)}_2>0$.
\end{lemma}
{\bf Proof}: 
From Lemma \ref{lem:13},
\begin{align}
\lambda^{1(2)}_1\sigma^{1(1)}_1&=\sum_{i=1}^4\lambda^{1(2)}_i\sigma^{1(1)}_i\nonumber\\
&=(Q^{1(2)}-Q^{1(1)},Q^0-Q^{1(1)}),\label{eqn:lamal3}\\
\lambda^{1(1)}_2\sigma^{1(2)}_2&=\sum_{i=1}^4\lambda^{1(1)}_i\sigma^{1(2)}_i\nonumber\\
&=(Q^{1(1)}-Q^{1(2)},Q^0-Q^{1(2)}),\label{eqn:lamal4}
\end{align}
so, by (\ref{eqn:lamal3})$+$(\ref{eqn:lamal4}) we obtain $\lambda^{1(2)}_1\sigma^{1(1)}_1+\lambda^{1(1)}_2\sigma^{1(2)}_2=(Q^{1(1)}-Q^{1(2)},Q^{1(1)}-Q^{1(2)})>0$.\hfill$\Box$

\bigskip

Now from Lemmas \ref{lem:13}, \ref{lem:14}, we see that $\lambda^{1(2)}_1<0$ or $\lambda^{1(1)}_2<0$ holds. So far, the points $P^1$ and $P^2$ have been treated exactly equally, hence without loss of generality, we assume
\begin{align}
\lambda^{1(2)}_1<0,\label{eqn:lambdanegative1}
\end{align}
and proceed to the next step. Here, we reconfirm the signs of the components of $\bm\lambda^{1(2)}$ as follows:
\begin{align}
\label{eqn:saikakunin}
\lambda^{1(2)}_1<0, \lambda^{1(2)}_2=0, \lambda^{1(2)}_3\geq0, \lambda^{1(2)}_4\geq0.
\end{align}

We make a classification based on the sings of components of $\bm\lambda^{1(1)}$. So far, the points $P^3$ and $P^4$ have been treated exactly equally, so without loss of generality, we have the following classification:

\medskip

Case 4-3-1: $\lambda^{1(1)}_1=0, \lambda^{1(1)}_2\geq0, \lambda^{1(1)}_3\geq0, \lambda^{1(1)}_4\geq0$

\medskip

Case 4-3-2: $\lambda^{1(1)}_1=0, \lambda^{1(1)}_2<0, \lambda^{1(1)}_3\geq0, \lambda^{1(1)}_4\geq0$

\medskip

Case 4-3-3: $\lambda^{1(1)}_1=0, \lambda^{1(1)}_2\geq0, \lambda^{1(1)}_3<0, \lambda^{1(1)}_4\geq0$

\medskip
\medskip
\noindent{\bf Case 4-3-1}\\
From Lemma \ref{lem:13}, we obtain $d(P^1,Q^{1(1)})<d(P^2,Q^{1(1)})=\cdots=d(P^4,Q^{1(1)})$ in a similar way to the proof of Lemma \ref{lem:8}. Therefore, by the Kuhn-Tucker condition (\ref{eqn:KT}), we have
\begin{theorem}
\label{theorem:9}
The center of \,$\Gamma$ is $Q^\ast=Q^{1(1)}$ and the radius is $d^\ast=d(P^2,Q^{1(1)})$.
\end{theorem}

\medskip
\noindent{\bf Case 4-3-2}\\
Define $Q^2\equiv\pi(Q^{1(1)}|L(P^3,P^4))=\pi(Q^0|L(P^3,P^4))$ and consider the inner products $\sigma^2_i=(P^i-Q^2,Q^0-Q^2),\,i=1,\cdots,4$. We have
\begin{lemma}
\label{lem:15}
$\sigma^2_i\left\{\begin{array}{ll}
<0, & i=1,2,\\
=0, & i=3,4.
\end{array}\right.$ 
\end{lemma}
{\bf Proof}: 
From $P^i-Q^2\perp Q^0-Q^2,\,i=3,4$, we have $\sigma^2_i=0,\,i=3,4$. Next, we will prove $\sigma^2_1<0$, $\sigma^2_2<0$. Note that $Q^2=\pi(Q^{1(2)}|L(P^3,P^4))=\pi(Q^0|L(P^3,P^4))$. From (\ref{eqn:null1}), 
\begin{align}
\lambda^{1(2)}_1\sigma^2_1&=\sum_{i=1}^4\lambda^{1(2)}_i\sigma^2_i\nonumber\\
&=(Q^{1(2)}-Q^2,Q^0-Q^2)\nonumber\\
&=(Q^{1(2)}-Q^2,Q^{1(2)}-Q^2)\nonumber\\
&>0,\nonumber
\end{align}
and by the assumption $\lambda^{1(2)}_1<0$, we obtain $\sigma^2_1<0$. Similarly, by considering $\sum_{i=1}^4\lambda^{1(1)}_i\sigma^2_i$, we obtain $\sigma^2_2<0$.\hfill$\Box$

\begin{lemma}
\label{lem:16}
$d(P^i,Q^2)<d(P^3,Q^2)=d(P^4,Q^2),\ i=1,2.$
\end{lemma}
{\bf Proof}: 
Similar to the proof of Lemma \ref{lem:8}.\hfill$\Box$
\begin{theorem}
\label{theorem:10}
The center of \,$\Gamma$ is $Q^\ast=Q^2$ and the radius is $d^\ast=d(P^3,Q^2)$.
\end{theorem}
{\bf Proof}: 
By Lemma \ref{lem:16} and the Kuhn-Tucker condition (\ref{eqn:KT}).\hfill$\Box$

\medskip
\noindent{\bf Case 4-3-3}\\
Put $Q^{1(3)}=\pi(Q^0|L(P^1,P^2,P^4))$, and consider the inner products $\sigma^{1(3)}_i=(P^i-Q^{1(3)},Q^0-Q^{1(3)}),\,
i=1,\cdots,4$. We have
\begin{lemma}
\label{lem:17}
$\sigma^{1(3)}_i\left\{\begin{array}{ll}
>0, & i=3,\\
=0, & i=1,2,4.
\end{array}\right.$ 
\end{lemma}
{\bf Proof}: 
From $P^i-Q^{1(3)}\perp Q^0-Q^{1(3)},\,i=1,2,4$, we have $\sigma^{1(3)}_i=0,\,i=1,2,4$. Thus, $\lambda^0_3\sigma^{1(3)}_3=\sum_{i=1}^4\lambda^0_i\sigma^{1(3)}_i=(Q^0-Q^{1(3)},Q^0-Q^{1(3)})>0$, so by the assumption $\lambda^0_3>0$, we obtain $\sigma^{1(3)}_3>0$.\hfill$\Box$

\begin{lemma}
\label{lem:18}
$\lambda^{1(3)}_1<0.$
\end{lemma}
{\bf Proof}: 
By Lemma \ref{lem:17}, we have
\begin{align}
\lambda^{1(1)}_3\sigma^{1(3)}_3&=\sum_{i=1}^4\lambda^{1(1)}_i\sigma^{1(3)}_i\nonumber\\
&=(Q^{1(1)}-Q^{1(3)},Q^0-Q^{1(3)}).\label{eqn:lamal5}
\end{align}
Further, by Lemma \ref{lem:13}, 
\begin{align}
\lambda^{1(3)}_1\sigma^{1(1)}_1&=\sum_{i=1}^4\lambda^{1(3)}_i\sigma^{1(1)}_i\nonumber\\
&=(Q^{1(3)}-Q^{1(1)},Q^0-Q^{1(1)}).\label{eqn:lamal6}
\end{align}
By (\ref{eqn:lamal5})$+$(\ref{eqn:lamal6}), we have
\begin{align}
\lambda^{1(1)}_3\sigma^{1(3)}_3+\lambda^{1(3)}_1\sigma^{1(1)}_1
&=(Q^{1(1)}-Q^{1(3)},Q^{1(1)}-Q^{1(3)})\nonumber\\
&>0.\label{eqn:lamal7}
\end{align}
By the assumption of Case 4-3-3, $\lambda^{1(1)}_3<0$ holds, by Lemma \ref{lem:17}, $\sigma^{1(3)}_3>0$ holds, and by Lemma \ref{lem:13}, $\sigma^{1(1)}_1<0$ holds, therefore, by (\ref{eqn:lamal7}), we obtain $\lambda^{1(3)}_1<0$.\hfill$\Box$

\bigskip

Here, we reconfirm the signs of the components of $\bm\lambda^{1(1)}$ and $\bm\lambda^{1(3)}$ as follows:
\begin{align}
&\lambda^{1(1)}_1=0, \lambda^{1(1)}_2\geq0, \lambda^{1(1)}_3<0, \lambda^{1(1)}_4\geq0,\label{eqn:sigcheck1}\\
&\lambda^{1(3)}_1<0, \lambda^{1(3)}_2\geq0, \lambda^{1(3)}_3=0, \lambda^{1(3)}_4\geq0.\label{eqn:sigcheck2}
\end{align}
If we exchange the points $P^2$ and $P^3$, then the second coordinate and the third coordinate are exchanged in (\ref{eqn:sigcheck1}) and (\ref{eqn:sigcheck2}), hence as a result, the signs of the components become the same combination as in Case 4-3-2 and (\ref{eqn:saikakunin}). In the proof of Case 4-3-2, the barycentric coordinate $\bm\lambda^0$ of $Q^0$ is not used, only $\bm\lambda^{1(1)}$ and $\bm\lambda^{1(3)}$ are used in the proof, therefore, exchanging $P^2$ and $P^3$ in the proof of Case 4-3-2 gives the proof of Case 4-3-3. Hence, in this Case 4-3-3, we have the following theorem.
\begin{theorem}
\label{theorem:11}
Put $Q^2{^\dagger}=\pi(Q^0|L(P^2,P^4))$. Then the center of \,$\Gamma$ is $Q^\ast=Q^2{^\dagger}$ and the radius is $d^\ast=d(P^2,Q^2{^\dagger})$.
\end{theorem}
\section{Problem of channel capacity}
\label{sec:channelcapacity}
In chapter \ref{sec:minimumenclosingcircle}, we obtained some theorems on the problem of smallest enclosing circle using Euclidean geometry. In particular, by using the distance, inner product, Pythagorean theorem and projection in the Euclidean space, we obtained a method of searching for the center $Q^\ast$ of the smallest enclosing circle by the projection algorithm. The problem of channel capacity has a similar geometric structure to that of smallest enclosing circle based on the similarity of (\ref{eqn:minmax2}) and (\ref{eqn:minimaxd}). 

In this chapter, we will consider the problem of channel capacity geometrically based on the results in chapter \ref{sec:minimumenclosingcircle}, to exploit a projection algorithm of searching for the output distribution that achieves the channel capacity.
\subsection{Information geometry}
The underlying geometry of the problem of channel capacity is the information geometry \cite{ama}, rather than the Euclidean geometry. A difference between the Euclidean geometry and the information geometry is that the Euclidean geometry uses one coordinate system but the information geometry uses two mutually dual coordinate systems. Amari \cite{ama} investigated $\alpha$-geometry for real $\alpha$, which is a family of geometric structures. The Euclidean geometry corresponds to $\alpha=0$ and the geometry of $\Delta^n$ corresponds to $\alpha=\pm1$, so, they can be regarded as a special case of $\alpha$-geometry. In $\alpha$-geometry, $\alpha$-divergence, inner product, Pythagorean theorem, $\alpha$-projection can be used. In the proof of theorems for the problem of smallest enclosing circle, we only used the Euclidean distance, inner product, Pythagorean theorem and projection among the properties of the Euclidean geometry. Thus, the resulting theorems or algorithms are expected to apply easily to the problem of channel capacity. In fact, we show it in the following.
\subsection{Geometric structure on $\Delta^n$}
Let $\Delta^n$ be the set of probability distributions with positive components on the output alphabet $\{y_1,\cdots,y_n\}$, i.e., 
\begin{align*}
\Delta^n=\{Q=(Q_1,\cdots,Q_n)|Q_j>0,j=1,\cdots,n,\sum_{j=1}^nQ_j=1\}.
\end{align*}
Geometric structure is introduced on $\Delta^n$ as follows \cite{ama}.
\subsubsection{Dual coordinate systems}
Two coordinate systems, in other words, two ways to specify $Q=(Q_1,\cdots,Q_n)\in\Delta^n$ are given on $\Delta^n$.
\begin{align*}
\eta\ &\mbox{\rm coordinate:}\\
&\bm\eta=(\eta_2,\cdots,\eta_n),\,\eta_j=Q_j,\,j=2,\cdots,n,\\[3mm]
\theta\ &\mbox{\rm coordinate:}\\
&\bm\theta=(\theta_2,\cdots,\theta_n),\,\theta_j=\log\ds\frac{Q_j}{Q_1},\,j=2,\cdots,n.
\end{align*}
The $\eta$ coordinate system and the $\theta$ coordinate system are mutually dual coordinate systems \cite{ama}.

\subsubsection{Geodesic}
A straight line with respect to the $\eta$ coordinate is called an $\eta$ geodesic. Let $\bm\eta^1,\bm\eta^2$ be the $\eta$ coordinates of $Q^1,Q^2\in\Delta^n$, respectively, then the $\eta$ geodesic passing through $Q^1,Q^2$ is defined by
\begin{align}
\bm\eta(t)=(1-t)\bm\eta^1+t\bm\eta^2\in\Delta^n,\,t\in\mathbb{R}.
\end{align}

Further, a straight line with respect to the $\theta$ coordinate is called a $\theta$ geodesic. Let $\bm\theta^1,\bm\theta^2$ be the $\theta$ coordinates of $Q^1,Q^2\in\Delta^n$, respectively, then the $\theta$ geodesic passing through $Q^1,Q^2$ is defined by
\begin{align}
\bm\theta(t)=(1-t)\bm\theta^1+t\bm\theta^2\in\Delta^n,\,t\in\mathbb{R}.
\end{align}
\subsection{Inner product, Pythagorean theorem and projection in $\Delta^n$}
\subsubsection{Inner product}
Consider three points $Q^1,Q^2,Q^3\in\Delta^n$. The inner product $(Q^1-Q^2,Q^3\ominus Q^2)$ is defined as follows. Let $\bm\eta^1=(\eta^1_2,\cdots,\eta^1_n),\bm\eta^2=(\eta^2_2,\cdots,\eta^2_n)$ be the $\eta$ coordinates of $Q^1,Q^2$, respectively, and $\bm\theta^2=(\theta^2_2,\cdots,\theta^2_n),\bm\theta^3=(\theta^3_2,\cdots,\theta^3_n)$ be the $\theta$ coordinates of $Q^2,Q^3$, respectively. Then the {\it inner product} is defined by 
\begin{align}
(Q^1-Q^2,Q^3\ominus Q^2)\equiv\sum_{j=2}^n(\eta^1_j-\eta^2_j)(\theta^3_j-\theta^2_j).
\end{align}
This is the inner product (in the usual sense) of the two tangent vectors $d\bm\eta(t)/dt|_{t=0}=\bm\eta^1-\bm\eta^2$ and $d\bm\theta(t)/dt|_{t=0}=\bm\theta^3-\bm\theta^2$ at $Q^2$ for two geodesics $\bm\eta(t)=(1-t)\bm\eta^2+t\bm\eta^1$ and $\bm\theta(t)=(1-t)\bm\theta^2+t\bm\theta^3$ passing through $Q^2$ \cite{ama}.

For $Q^1,Q^2,Q^3\in\Delta^n$, the $\eta$ geodesic $\bm\eta(t)=(1-t)\bm\eta^2+t\bm\eta^1$ and the $\theta$ geodesic $\bm\theta(t)=(1-t)\bm\theta^2+t\bm\theta^3$ are said to be {\it orthogonal} at $Q^2$ if \cite{ama}
\begin{align}
(Q^1-Q^2,Q^3\ominus Q^2)=0.
\end{align}
We have the following lemmas.
\begin{lemma}
\label{lem:20}
For $P^i\,(i=1,\cdots,m),Q,R\in\Delta^n$, consider the inner products $\sigma_i=(P^i-Q,R\ominus Q),\,i=1,\cdots,m$. If $\sum_{i=1}^m\lambda_i=1$, then 
\begin{align}
\sum_{i=1}^m\lambda_i\sigma_i=\left(\sum_{i=1}^m\lambda_iP^i-Q,R\ominus Q\right).
\end{align}
\end{lemma}
{\bf Proof}: 
By the definition of inner product and a simple calculation.\hfill$\Box$

\begin{lemma}
\label{lem:21}
For any $P,Q,R\in\Delta^n$, we have
\begin{align*}
(P-Q,R\ominus Q)=-(Q-P,R\ominus P)+(P-Q,P\ominus Q).
\end{align*}
\end{lemma}
{\bf Proof}: 
By a simple calculation.\hfill$\Box$

\begin{lemma}
\label{lem:22}
For any $P,Q,R\in\Delta^n$, we have
\begin{align*}
(P-Q,R\ominus Q)=D(P\|Q)+D(Q\|R)-D(P\|R).
\end{align*}
\end{lemma}
{\bf Proof}: 
By a simple calculation (see also \cite{ama}).\hfill$\Box$

\subsubsection{Pythagorean theorem}
For three points $P,Q,R$ in $\Delta^n$, the following Pythagorean theorem and its inequality version hold.
\begin{theorem}(Pythagorean)
\label{theorem:12}
For $P,Q,R\in\Delta^n$, we have
%
%
%
\begin{align}
(P-Q,R\ominus Q)\gtreqqless0 \iff D(P\|Q)+D(Q\|R)\gtreqqless D(P\|R).
\end{align}
\end{theorem}
{\bf Proof}: 
By Lemma \ref{lem:22} (see also \cite{ama}).\hfill$\Box$

\subsubsection{Projection by Kullback-Leibler divergence}
For $P^1,\cdots,$ $P^m\in\Delta^n$, the affine subspace $L(P^1,\cdots,P^m)\subset\Delta^n$ spanned by $P^1,\cdots,P^m$ is defined by 
\begin{align}
\label{eqn:affinesubspaceC}
L(P^1,\cdots,P^m)=\left\{\sum_{i=1}^m\lambda_iP^i\,\Big|\sum_{i=1}^m\lambda_i=1\right\}\cap\Delta^n.
\end{align}

\begin{figure}[t]
\begin{center}
\begin{overpic}[width=7.5cm]{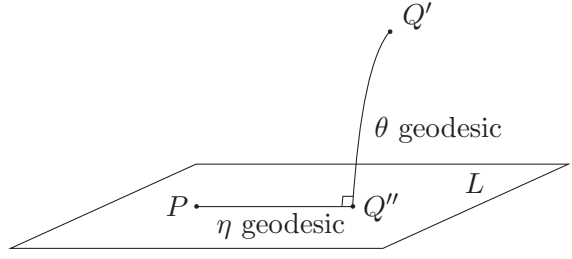}
\put(70,40){$Q'$}
\put(63,6){$Q''$}
\put(28,6){$P$}
\put(81,10){$L$}
\put(65,20){$\theta$ geodesic}
\put(37,3){$\eta$ geodesic}
\end{overpic}
\end{center}
\caption{Projection $Q''=\pi(Q'|L)$ of $Q'$ onto affine subspace $L$}
\label{fig:4}
\end{figure}

For $Q'\in\Delta^n$ and a subset $L\subset\Delta^n$, the $Q=Q''$ that achieves $\min_{Q\in L}D(Q\|Q')$ is called the {\it projection} of $Q'$ onto $L$, and denoted by $Q''=\pi(Q'|L)$. In this paper, we consider only affine subspaces as $L$.
\begin{lemma}
\label{lem:23}
Let $L$ be an affine subspace in $\Delta^n$. Then, for any $Q'\in\Delta^n$, the projection $Q''=\pi(Q'|L)$ exists and is unique. Moreover, $Q''=\pi(Q'|L)$ is equivalent to that $(P-Q'',Q'\ominus Q'')=0$ holds for any $P\in L$ (see Fig.\ref{fig:4}).
\end{lemma}
{\bf Proof}: 
see \cite{ama}.\hfill$\Box$

\subsection{Equidistant point and projection of equidistant point}
We will calculate the channel capacity $C$ and the output distribution $Q^\ast$ that achieves $C$, in case that the row vectors $P^1,\cdots,P^m$ of the channel matrix $\Phi$ in (\ref{eqn:thechannelmatrix}) are in general position. In chapter \ref{sec:minimumenclosingcircle}, we investigated the problem of smallest enclosing circle based on the Euclidean geometry in case that the points $P^1,\cdots,P^m$ are in general position. In this chapter, we will consider the problem of channel capacity based on the information geometry.
\subsubsection{Equidistant point from $P^1,\cdots,P^m$}
For a channel matrix $\Phi$ in (\ref{eqn:thechannelmatrix}), a matrix $\Psi\in\mathbb{R}^{(m-1)\times n}$ is defined by
\begin{align}
\Psi&=\left(\begin{array}{c}
P^2-P^1\\
\vdots\\
P^m-P^1
\end{array}\right)\nonumber\\
&=\left(\begin{array}{ccc}
P^2_1-P^1_1 & \cdots & P^2_n-P^1_n\\
\vdots & & \vdots\\
P^m_1-P^1_1 & \cdots & P^m_n-P^1_n
\end{array}\right).
\end{align}
Then, $P^1,\cdots,P^m$ are said to be {\it in general position}, if the vectors $P^2-P^1,\cdots,P^m-P^1$ are linearly independent, or
\begin{align}
{\rm rank}\,\Psi=m-1.\label{eqn:rankA}
\end{align}

Now, we assume in this chapter that $P^1,\cdots,P^m$ are in general position. Then, denote by $\Psi'\in\mathbb{R}^{(m-1)\times(n-1)}$ the matrix that is made by removing the first column of $\Psi$, i.e., 
\begin{align}
\Psi'=\left(\begin{array}{ccc}
P^2_2-P^1_2 & \cdots & P^2_n-P^1_n\\
\vdots & & \vdots\\
P^m_2-P^1_2 & \cdots & P^m_n-P^1_n
\end{array}\right).
\end{align}
We see rank$\,\Psi'=m-1$. In fact, let $P^i{'}=(P^i_2,\cdots,P^i_n),\,i=1,\cdots,m$, and suppose $\sum_{i=2}^mc_i(P^i{'}-P^1{'})=O$. Noticing $\sum_{j=1}^nP^i_j=1,\,i=1,\cdots,m$, we have $\sum_{i=2}^mc_i(P^i-P^1)=O$, therefore, from (\ref{eqn:rankA}), $c_i=0,\,i=2,\cdots,m$.

Now, similarly to the problem of smallest enclosing circle, we consider an output distribution that has the equal Kullback-Leibler divergence from $P^1,\cdots,P^m$, i.e., we consider $Q\in\Delta^n$ that satisfies
\begin{align}
D(P^i\|Q)=D(P^1\|Q),\,i=2,\cdots,m.\label{eqn:toudivergence1}
\end{align}
Let $\bm\theta=(\theta_2,\cdots,\theta_n)$ be the $\theta$ coordinate of $Q$, and $H(P^i)=-\sum_{j=1}^nP^i_j\log P^i_j$ be the entropy of $P^i,\,i=1,\cdots,m$. Then by a simple calculation, we have from (\ref{eqn:toudivergence1}),
\begin{align}
D(P^i&\|Q)-D(P^1\|Q)\nonumber\\
&=-\ds\sum_{j=2}^n(P^i_j-P^1_j)\,\theta_j-H(P^i)+H(P^1)\nonumber\\
&=0,\,i=2,\cdots,m.\label{eqn:toudivergence2}
\end{align}
Putting $\bm{b}=(-H(P^2)+H(P^1),\cdots,-H(P^m)+H(P^1))\in\mathbb{R}^{m-1}$, we can rewrite (\ref{eqn:toudivergence2}) as
\begin{align}
\Psi'\hspace{0.8mm}{^t}\hspace{-0.1mm}\bm\theta=\hspace{0.3mm}{^t}\hspace{-0.3mm}\bm{b}.\label{eqn:toudivergence3}
\end{align}
Because rank$\,\Psi'=m-1$, the equation (\ref{eqn:toudivergence3}) has a solution $\bm\theta$, but it is not necessarily unique. From a solution $\bm\theta=(\theta_2,\cdots,\theta_n)$ of (\ref{eqn:toudivergence3}), we make $Q=(Q_1,\cdots,Q_n)\in\Delta^n$ by
\begin{align}
\label{eqn:toudivergence4}
\left\{\begin{array}{l}
Q_1=\left(1+\ds\sum_{j=2}^n\exp\theta_j\right)^{-1},\\[6mm]
Q_j=Q_1\exp\theta_j,\,j=2,\cdots,n.
\end{array}
\right.
\end{align}
Then the $\theta$ coordinate of $Q$ becomes $\bm\theta$.
\begin{lemma}
\label{lem:25}
Let $L_0=L(P^1,\cdots,P^m)$ be the affine subspace spanned by $P^1,\cdots,P^m$ (see (\ref{eqn:affinesubspaceC})). Make $Q\in\Delta^n$ by $(\ref{eqn:toudivergence4})$ from a solution $\bm\theta$ of the equation $(\ref{eqn:toudivergence3})$ and put $Q^0=\pi(Q|L_0)$. Then, $Q^0$ is the unique point in $L_0$ with
\begin{align}
\label{eqn:equalKLdivergence}
D(P^i\|Q^0)=D(P^1\|Q^0),\,i=2,\cdots,m.
\end{align}
\end{lemma}
{\bf Proof}: 
For $Q,\,Q'\in\Delta^n$ with (\ref{eqn:toudivergence3}), (\ref{eqn:toudivergence4}), let $Q^0=\pi(Q|L_0)$, ${Q^0}{'}=\pi(Q'|L_0)$. We will prove $Q^0={Q^0}{'}$. Put $Q^0=(Q^0_1,\cdots,Q^0_n),\,Q^0{'}=(Q^0_1{'},\cdots,Q^0_n{'})$. Both $Q$ and $Q'$ satisfy the equation (\ref{eqn:toudivergence1}), so from Theorem \ref{theorem:12} and Lemma \ref{lem:23}, we have
\begin{align}
&\ds\sum_{j=1}^nP^i_j\log\ds\frac{P^i_j}{Q^0_j}
=\ds\sum_{j=1}^nP^1_j\log\ds\frac{P^1_j}{Q^0_j},\ i=2,\cdots,m,\label{eqn:pythagoras1}\\
&\ds\sum_{j=1}^nP^i_j\log\ds\frac{P^i_j}{Q^0_j{'}}
=\ds\sum_{j=1}^nP^1_j\log\ds\frac{P^1_j}{Q^0_j{'}},\ i=2,\cdots,m.\label{eqn:pythagoras2}
\end{align}
Subtracting (\ref{eqn:pythagoras1}) from (\ref{eqn:pythagoras2}),
\begin{align}
\ds\sum_{j=1}^nP^i_j\log\ds\frac{Q^0_j}{Q^0_j{'}}
=\ds\sum_{j=1}^nP^1_j\log\ds\frac{Q^0_j}{Q^0_j{'}},\ i=2,\cdots,m.\label{eqn:pythagoras3}
\end{align}
From (\ref{eqn:pythagoras3}), we can write
\begin{align}
\ds\sum_{j=1}^nP^i_j\log\ds\frac{Q^0_j}{Q^0_j{'}}=\kappa\ \mbox{\rm (constant)},\ i=1,\cdots,m.\label{eqn:pythagoras4}
\end{align}
Let $\bm\lambda^0=(\lambda^0_1,\cdots,\lambda^0_m)$, $\bm\lambda^0{'}=(\lambda^0_1{'},\cdots,\lambda^0_m\!{'})$ be the barycentric coordinates of $Q^0,\,Q^0{'}\in L_0$ about $P^1,\cdots,P^m$, respectively. Then, from (\ref{eqn:pythagoras4}),
\begin{align}
\kappa&=\ds\sum_{i=1}^m\lambda^0_i\ds\sum_{j=1}^nP^i_j\log\ds\frac{Q^0_j}{Q^0_j{'}}
=D(Q^0\|Q^0{'}),\label{eqn:pythagoras5}\\
\kappa&=\ds\sum_{i=1}^m\lambda^0_i{'}\ds\sum_{j=1}^nP^i_j\log\ds\frac{Q^0_j}{Q^0_j{'}}
=-D(Q^0{'}\|Q^0).\label{eqn:pythagoras6}
\end{align}
Subtracting (\ref{eqn:pythagoras6}) from (\ref{eqn:pythagoras5}), 
\begin{align}
0=D(Q^0\|Q^0{'})+D(Q^0{'}\|Q^0),
\end{align}
thus, we obtain $Q^0=Q^0{'}$.\hfill$\Box$

\medskip

$Q^0\in L_0=L(P^1,\cdots,P^m)$ with (\ref{eqn:equalKLdivergence}) is called the {\it equidistant point} from $P^1,\cdots,P^m$. The existence and uniqueness of $Q^0$ is guaranteed by Lemma \ref{lem:25}. It might be better to call $Q^0$ an equi-divergence distribution, but, by analogy with smallest enclosing circle, we call it an equidistant point, too.

\subsubsection{Projection of equidistant point}
Now, we further define $L_k=L(P^{k+1},\cdots,P^m),\,k=0,1,\cdots,m-2$. $L_0\supset L_1\supset\cdots\supset L_k\supset\cdots$ is a decreasing sequence of affine subspaces whose dimensions are decreasing by $1$.

Let $Q^1=\pi(Q^0|L_1)$ denote the projection of $Q^0$ onto $L_1$. Further, we define $Q^k=\pi(Q^{k-1}|L_k),$ $k=1,\cdots,m-2$.
\begin{lemma}
\label{lem:26}
$Q^k=\pi(Q^0|L_k),\,k=0,1,\cdots,m-2.$
\end{lemma}
{\bf Proof}: 
It is trivial for $k=0,1$ by definition. Next for $k=2$, putting $Q^2{'}=\pi(Q^0|L_2)$, we will prove $Q^2{'}=Q^2\equiv\pi(Q^1|L_2)$. Since $Q^1=\pi(Q^0|L_1)$, for any $Q\in L_1$ we have by Theorem \ref{theorem:12} and Lemma \ref{lem:23}
\begin{align}
D(Q\|Q^0)=D(Q\|Q^1)+D(Q^1\|Q^0).\label{eqn:pythagoras7}
\end{align}
Therefore, by (\ref{eqn:pythagoras7}), with respect to $Q\in L_2(\subset L_1)$ minimizing $D(Q\|Q^0)$ and minimizing $D(Q\|Q^1)$ are equivalent. Because the projections $\pi(Q^1|L_2)$ and $\pi(Q^0|L_2)$ are unique by Lemma \ref{lem:23}, we obtain $Q^2{'}=Q^2$. For $k\geq3$, it is proved by mathematical induction.\hfill$\Box$

\begin{lemma}
\label{lem:27}
$(Q^i-Q^k,Q^0\ominus Q^k)=(Q^i-Q^k,Q^i\ominus Q^k),\,i=0,1,\cdots,k,k=0,1,\cdots,m-2.$
\end{lemma}
{\bf Proof}: 
By Lemma \ref{lem:21}, we have
\begin{align}
&(Q^i-Q^k,Q^0\ominus Q^k)\nonumber\\
&=-(Q^k-Q^i,Q^0\ominus Q^i)+(Q^i-Q^k,Q^i\ominus Q^k).
\end{align}
By Lemma \ref{lem:26}, $Q^i=\pi(Q^0|L_i)$, and by $k\geq i$, $Q^k\in L_i$, so we obtain $(Q^k-Q^i,Q^0\ominus Q^i)=0$ by Lemma \ref{lem:23}.\hfill$\Box$

\begin{lemma}
\label{lem:28}
For any $P,Q\in\Delta^n,\,P\neq Q$, we have $(P-Q,P\ominus Q)>0$.
\end{lemma}
{\bf Proof}: 
By Lemma \ref{lem:22}, we obtain $(P-Q,P\ominus Q)=D(P\|Q)+D(Q\|P)>0$.\hfill$\Box$

\subsection{Search for $Q^\ast$ by projection algorithm}
For a given channel matrix $\Phi$ in (\ref{eqn:thechannelmatrix}), let $C$ be the channel capacity and $Q^\ast$ be the output distribution that achieves $C$. In this section, similarly to section \ref{subsec:tansaku1}, we will show that $Q^\ast$ and $C$ are obtained by the projection algorithm, under the assumption that the row vectors of the channel matrix $\Phi$ are in general position and in the situations 1 and 2 below.
\subsection{Situation 1 $[$There is just one negative component of barycentric coordinate at every projection.$]$}
\label{subsec:channeljoukyou1}
We consider the following assumption similar to section \ref{subsec:joukyou1}.

\noindent\underline{Assumption of situation 1}\ \ 
Assume $P^1,\cdots,P^m\in\Delta^n$ are in general position, and let $L_k=L(P^{k+1},\cdots,P^m),\,k=0,1,\cdots,$ $m-2$ be the affine subspace spanned by $P^{k+1},\cdots,P^m$. Let $Q^0\in L_0$ be the equidistant point from $P^1,\cdots,P^m$, and define $Q^k=\pi(Q^{k-1}|L_k),\,k=1,\cdots,m-2$. The barycentric coordinate of $Q^k$ about $P^1,\cdots,P^m$ is denoted by $\bm\lambda^k=(\lambda^k_1,\cdots,\lambda^k_m)$. Let $K=0,1,\cdots,m-2$. We assume that for $k=0,1,\cdots,K-1$, there is just one negative component of $\bm\lambda^k$, and for $k=K$, all the components of $\bm\lambda^K$ are non-negative. That is, for $k=0,1,\cdots,K-1$, 
\begin{subnumcases}
{\lambda^k_i}
=0, & \hspace{-4mm}$i=1,\cdots,k,$\label{eqn:3a}\\[-1mm]
<0, & \hspace{-4mm}$i=k+1,$\label{eqn:3b}\\[-1mm]
>0, & \hspace{-4mm}$i=k+2,\cdots,m,$\label{eqn:3c}
\end{subnumcases}
and for $k=K$,
\begin{subnumcases}
{\lambda^K_i}
=0, & \hspace{-4mm}$i=1,\cdots,K,$\label{eqn:4a}\\[-1mm]
>0, & \hspace{-4mm}$i=K+1,\cdots,m$.\label{eqn:4b}
\end{subnumcases}

\noindent If $K=0$, we assume $\lambda^0_i\geq0,\,i=1,\cdots,m$. 

\noindent(The end of Assumption of situation 1)

\medskip

Here, we consider the inner products $\sigma_i=(P^i-Q^K,Q^0\ominus Q^K),\,i=1,\cdots,m$.
\begin{lemma}
\label{lem:29}
$\sigma_i\left\{\begin{array}{ll}
<0, & i=1,\cdots,K,\\
=0, & i=K+1,\cdots,m.
\end{array}\right.$
\end{lemma}
{\bf Proof}: 
By Lemma \ref{lem:26}, $Q^K=\pi(Q^0|L_K)$, and since $P^{K+1}\cdots,P^m\in L_K$, we have, by Lemma \ref{lem:23}, 
\begin{align}
\sigma_i=0,\,i=K+1,\cdots,m.\label{eqn:sigma_i3}
\end{align}
Next, we will prove $\sigma_i<0,\,i=1,\cdots,K$ by mathematical induction in the order of $i=K,K-1,\cdots,1$.

\noindent(I) Prove $\sigma_K<0$:

By (\ref{eqn:3a}), (\ref{eqn:sigma_i3}), we have
\begin{align}
\lambda^{K-1}_K\sigma_K&=\sum_{i=1}^m\lambda^{K-1}_i\sigma_i\nonumber\\
&=(Q^{K-1}-Q^K,Q^0\ominus Q^K)\nonumber\\
&=(Q^{K-1}-Q^K,Q^{K-1}\ominus Q^K)\ \ {\rm (by\ Lemma\ \ref{lem:27})}\nonumber\\
&>0\ \ {\rm (by\ Lemma\ \ref{lem:28})}.\label{eqn:lamal16}
\end{align}
By (\ref{eqn:3b}), $\lambda^{K-1}_K<0$, thus we obtain $\sigma_K<0$ by (\ref{eqn:lamal16}).

\noindent(II) Assuming $\sigma_K<0,\,\sigma_{K-1}<0,\cdots,\sigma_{k+1}<0$, prove $\sigma_k<0$:

By (\ref{eqn:3a}), (\ref{eqn:sigma_i3}), we have
\begin{align}
\lambda^{k-1}_k\sigma_k+\sum_{i=k+1}^K\lambda^{k-1}_i\sigma_i&=\sum_{i=1}^m\lambda^{k-1}_i\sigma_i\nonumber\\
&=(Q^{k-1}-Q^K,Q^0\ominus Q^K)\nonumber\\
&=(Q^{k-1}-Q^K,Q^{k-1}\ominus Q^K)\nonumber\\
&>0.\label{eqn:lamal17}
\end{align}
By the induction hypothesis, $\sigma_i<0,\,i=k+1,\cdots,K$, and by (\ref{eqn:3b}), $\lambda^{k-1}_k<0$, by (\ref{eqn:3c}), $\lambda^{k-1}_i>0,\,i=k+1,\cdots,K$, thus by (\ref{eqn:lamal17}), we obtain $\sigma_k<0$.

\noindent(III) By the above (I), (II), we have $\sigma_i<0,\,i=1,\cdots,K$.\hfill$\Box$

\begin{lemma}
\label{lem:30}
$D(P^i\|Q^K)<D(P^{K+1}\|Q^K)=\cdots=D(P^m\|Q^K),\,i=1,\cdots,K.$
\end{lemma}
{\bf Proof}: 
By Lemma \ref{lem:29} and Theorem \ref{theorem:12}.\hfill$\Box$

\begin{theorem}
\label{theorem:13}
The output distribution that achieves the channel capacity is $Q^\ast=Q^K$, and the channel capacity is $C=D(P^{K+1}\|Q^K)$.
\end{theorem}
{\bf Proof}: 
By (\ref{eqn:4a}), (\ref{eqn:4b}), Lemma \ref{lem:30} and the Kuhn-Tucker condition (\ref{eqn:KT2}).\hfill$\Box$

\bigskip

We see that the above Lemmas \ref{lem:29},\,\ref{lem:30}, Theorem \ref{theorem:13} and their proofs are very similar to Lemmas \ref{lem:7},\,\ref{lem:8}, Theorem \ref{theorem:2} and their proofs in the problem of smallest enclosing circle in chapter \ref{sec:minimumenclosingcircle}. This is because that the problems of smallest enclosing circle and channel capacity can be solved using only common properties of $\alpha$-geometry.
\subsection{Situation 2 $[$$m=2,3,4$ and $n$ is arbitrary.$]$}
We will calculate by the projection algorithm the capacity achieving output distribution $Q^\ast$ and the channel capacity $C$, under the assumption that the row vectors $P^1,\cdots,P^m$ of the channel matrix $\Phi$ in (\ref{eqn:thechannelmatrix}) are in general position in $\Delta^n$.

Similarly to chapter \ref{sec:minimumenclosingcircle}, our goal is to find an algorithm to calculate $Q^\ast$ and $C$ for every $m$, but at present it is possible to solve only for $m=2,3,4$.

As can be seen from the results of the previous section, the proofs of lemmas and theorems in the case of smallest enclosing circle are almost the same as those of channel capacity. Because those are similar also in this section, the proofs of the following lemmas and theorems will be described simply.
\subsubsection{Case of $m=2$}
Consider the channel capacity of a channel matrix $\Phi$ with two different row vectors $P^1,P^2\in\Delta^n$. Let $L_0=L(P^1,P^2)$, and $Q^0\in L_0$ be the equidistant point from $P^1,P^2$, i.e.,
\begin{align}
D(P^1\|Q^0)=D(P^2\|Q^0),\,Q^0\in L_0.\label{eqn:toudivergence5}
\end{align}
Let $\bm\lambda^0=(\lambda^0_1,\lambda^0_2)$ be the barycentric coordinate of $Q^0$ about $P^1,P^2$. Because $Q^0$ is between $P^1$ and $P^2$ in $L_0$, we have $\lambda^0_1>0,\lambda^0_2>0$. Thus, by (\ref{eqn:toudivergence5}) and the Kuhn-Tucker condition (\ref{eqn:KT2}), we have
\begin{theorem}
\label{theorem:14}
The output distribution that achieves the channel capacity is $Q^\ast=Q^0$, and the channel capacity is $C=D(P^1\|Q^0)$.
\end{theorem}
\subsubsection{Case of $m=3$}
Suppose the channel matrix consists of three row vectors $P^1,P^2,P^3\in\Delta^n$ in general position. Let $L_0=L(P^1,P^2,P^3)$ and let $Q^0\in L_0$ be the equidistant point from $P^1,P^2,P^3$. Let $\bm\lambda^0=(\lambda^0_1,\lambda^0_2,\lambda^0_3)$ be the barycentric coordinate of $Q^0$ about $P^1,P^2,P^3$. 

Based on the signs of the components of $\bm\lambda^0$, we have the following classification without loss of generality.

\medskip

Case 3-1: $\lambda^0_1\geq0,\lambda^0_2\geq0,\lambda^0_3\geq0$

\medskip

Case 3-2: $\lambda^0_1<0,\lambda^0_2\geq0,\lambda^0_3\geq0$

\medskip

As in the problem of smallest enclosing circle, all the cases are exhausted by Case 3-1 and Case 3-2.

\medskip
\noindent{\bf Case 3-1}\\
By the Kuhn-Tucker condition (\ref{eqn:KT2}), we have
\begin{theorem}
\label{theorem:15}
The output distribution that achieves the channel capacity is $Q^\ast=Q^0$ and the channel capacity is $C=D(P^1\|Q^0)$.
\end{theorem}

\medskip

\noindent{\bf Case 3-2}\\
Similarly to the Case 3-2 in chapter \ref{sec:minimumenclosingcircle}, this case is the situation 1, hence we have
\begin{theorem}
\label{theorem:16}
The output distribution that achieves the channel capacity is $Q^\ast=Q^1$ and the channel capacity is $C=D(P^2\|Q^1)$.
\end{theorem}
\subsubsection{Case of $m=4$}
Suppose the channel matrix $\Phi$ consists of four row vectors $P^1,P^2,P^3,P^4\in\Delta^n$ in general position. Let $L_0=L(P^1,\cdots,P^4)$ and $Q^0\in L_0$ be the equidistant point from $P^1,\cdots,P^4$. Let $\bm\lambda^0=(\lambda^0_1,\cdots,\lambda^0_4)$ be the barycentric coordinate of $Q^0$ about $P^1,\cdots,P^4$.

Without loss of generality, we have the following classification:

\medskip

Case 4-1: $\lambda^0_1\geq0,\lambda^0_2\geq0, \lambda^0_3\geq0, \lambda^0_4\geq0$

\medskip

Case 4-2: $\lambda^0_1<0,\lambda^0_2\geq0, \lambda^0_3\geq0, \lambda^0_4\geq0$

\medskip

Case 4-3: $\lambda^0_1<0,\lambda^0_2<0, \lambda^0_3\geq0, \lambda^0_4\geq0$

\medskip
\medskip
\noindent{\bf Case 4-1}\\
By the Kuhn-Tucker condition (\ref{eqn:KT2}), we have
\begin{theorem}
\label{theorem:17}
The output distribution that achieves the channel capacity is $Q^\ast=Q^0$ and the channel capacity is $C=D(P^1\|Q^0)$.
\end{theorem}

\medskip

\noindent{\bf Case 4-2}\\
Let $L_1=L(P^2,P^3,P^4)$ and $Q^1=\pi(Q^0|L_1)$, and denote by $\bm\lambda^1=(\lambda^1_1,\cdots,\lambda^1_4)$ the barycentric coordinate of $Q^1$ about $P^1,\cdots,P^4$. Since $P^1\notin L_1$, we have $\lambda^1_1=0$.

Based on the signs of the components of $\bm\lambda^1$, we have the following classification without loss of generality.

\medskip

Case 4-2-1: $\lambda^1_1=0,\lambda^1_2\geq0, \lambda^1_3\geq0, \lambda^1_4\geq0$

\medskip

Case 4-2-2: $\lambda^1_1=0,\lambda^1_2<0, \lambda^1_3\geq0, \lambda^1_4\geq0$

\medskip

\noindent Consider the inner products $\sigma_i=(P^i-Q^1,Q^0\ominus Q^1),\,i=1,\cdots,4$.

\medskip
\noindent{\bf Case 4-2-1}\\
This case is the situation 1, hence we have
\begin{theorem}
\label{theorem:18}
The output distribution that achieves the channel capacity is $Q^\ast=Q^1$ and the channel capacity is $C=D(P^2\|Q^1)$.
\end{theorem}

\medskip

\noindent{\bf Case 4-2-2}\\
Let us define $Q^2=\pi(Q^1|L_2)$ with $L_2=L(P^3,P^4)$, then $Q^2$ is between $P^3$ and $P^4$ in the line $L_2$. Denoting by $\bm\lambda^2=(\lambda^2_1,\cdots,\lambda^2_4)$ the barycentric coordinate of $Q^2$ about $P^1,\cdots,P^4$, we have $\lambda^2_1=\lambda^2_2=0, \lambda^2_3>0, \lambda^2_4>0$, hence this is the situation 1. So, we have
\begin{theorem}
\label{theorem:19}
Let $Q^2=\pi(Q^1|L(P^3,P^4))$. 
The output distribution that achieves the channel capacity is $Q^\ast=Q^2$ and the channel capacity is $C=D(P^3\|Q^2)$.
\end{theorem}

\medskip

This completes Case 4-2, then let us consider Case 4-3.
\medskip

\noindent{\bf Case 4-3}\\
Here, let us define $Q^{1(1)}=\pi(Q^0|L(P^2,P^3,P^4))$ and $Q^{1(2)}=\pi(Q^0|L(P^1,P^3,P^4))$. Then denote by $\bm\lambda^{1(1)}=(\lambda^{1(1)}_1,\cdots,\lambda^{1(1)}_4)$ and $\bm\lambda^{1(2)}=(\lambda^{1(2)}_1,\cdots,\lambda^{1(2)}_4)$ the barycentric coordinates of $Q^{1(1)}$ and $Q^{1(2)}$ about $P^1,\cdots,P^4$, respectively. Consider the inner products 
\begin{align}
&\sigma^{1(1)}_i=(P^i-Q^{1(1)},Q^0\ominus Q^{1(1)}),\ i=1,\cdots,4,\\
&\sigma^{1(2)}_i=(P^i-Q^{1(2)},Q^0\ominus Q^{1(2)}),\ i=1,\cdots,4.
\end{align}
\begin{lemma}
\label{lem:35}

\ \ $\sigma^{1(1)}_i\left\{\begin{array}{ll}
<0, & i=1,\\
=0, & i=2,3,4,
\end{array}\right.$
$\sigma^{1(2)}_i\left\{\begin{array}{ll}
<0, & i=2,\\
=0, & i=1,3,4.
\end{array}\right.$
\end{lemma}
\begin{lemma}
\label{lem:36}
$\lambda^{1(2)}_1\sigma^{1(1)}_1+\lambda^{1(1)}_2\sigma^{1(2)}_2>0.$
\end{lemma}
The above Lemmas \ref{lem:35},\,\ref{lem:36} are proved in the same way as Lemmas \ref{lem:13},\,\ref{lem:14}.

\bigskip

Now from Lemmas \ref{lem:35},\,\ref{lem:36}, we see that $\lambda^{1(2)}_1<0$ or $\lambda^{1(1)}_2<0$ holds. So far, the points $P^1$ and $P^2$ have been treated exactly equally, hence without loss of generality, we assume $\lambda^{1(2)}_1<0$. Based on the signs of the components of $\bm\lambda^{1(1)}$, we have the following classification:

\medskip

Case 4-3-1: $\lambda^{1(1)}_1=0, \lambda^{1(1)}_2\geq0, \lambda^{1(1)}_3\geq0, \lambda^{1(1)}_4\geq0$

\medskip

Case 4-3-2: $\lambda^{1(1)}_1=0, \lambda^{1(1)}_2<0, \lambda^{1(1)}_3\geq0, \lambda^{1(1)}_4\geq0$

\medskip

Case 4-3-3: $\lambda^{1(1)}_1=0, \lambda^{1(1)}_2\geq0, \lambda^{1(1)}_3<0, \lambda^{1(1)}_4\geq0$

\medskip
\medskip
\noindent{\bf Case 4-3-1}
\begin{theorem}
\label{theorem:20}
The output distribution that achieves the channel capacity is $Q^\ast=Q^{1(1)}$ and the channel capacity is $C=D(P^2\|Q^{1(1)})$.
\end{theorem}
{\bf Proof}: 
Similar to the proof of Theorem \ref{theorem:9}.\hfill$\Box$

\medskip
\noindent{\bf Case 4-3-2}\\
Define $Q^2\equiv\pi(Q^{1(1)}|L(P^3,P^4))(=\pi(Q^0|L(P^3,P^4)))$, and consider the inner products $\sigma^2_i=(P^i-Q^2,Q^0\ominus Q^2),\,i=1,\cdots,4$.
\begin{lemma}
\label{lem:37}
$\sigma^2_i\left\{\begin{array}{ll}
<0, & i=1,2,\\
=0, & i=3,4.
\end{array}\right.$
\end{lemma}
\begin{lemma}
\label{lem:38}
$D(P^i\|Q^2)<D(P^3\|Q^2)=D(P^4\|Q^2),\,i=1,2.$
\end{lemma}
\begin{theorem}
\label{theorem:21}
The output distribution that achieves the channel capacity is $Q^\ast=Q^2$ and the channel capacity is $C=D(P^3\|Q^2)$.
\end{theorem}
The above Lemmas \ref{lem:37},\,\ref{lem:38} and Theorem \ref{theorem:21} are proved in the same way as Lemmas \ref{lem:15},\,\ref{lem:16} and Theorem \ref{theorem:10}.

\medskip
\noindent{\bf Case 4-3-3}\\
Let us define $Q^{1(3)}=\pi(Q^0|L(P^1,P^2,P^4))$, and denote by $\bm\lambda^{1(3)}=(\lambda^{1(3)}_1,\cdots,\lambda^{1(3)}_4)$ the barycentric coordinate of $Q^{1(3)}$ about $P^1,\cdots,P^4$. Consider the inner products $\sigma^{1(3)}_i=(P^i-Q^{1(3)},Q^0\ominus Q^{1(3)}),\,i=1,\cdots,4$.
\begin{lemma}
\label{lem:39}
$\sigma^{1(3)}_i\left\{\begin{array}{ll}
>0, & i=3,\\
=0, & i=1,2,4.
\end{array}\right.$
\end{lemma}
\begin{lemma}
\label{lem:40}
$\lambda^{1(3)}_1<0.$
\end{lemma}
\begin{theorem}
\label{theorem:22}
Let $Q^2{^\dagger}\equiv\pi(Q^0|L(P^2,P^4))$. 
The output distribution that achieves the channel capacity is $Q^\ast=Q^2{^\dagger}$ and the channel capacity is $C=D(P^2\|Q^2{^\dagger})$.
\end{theorem}
The above Lemmas \ref{lem:39},\,\ref{lem:40} and Theorem \ref{theorem:22} are proved in the same way as Lemmas \ref{lem:17},\,\ref{lem:18} and Theorem \ref{theorem:11}.

Summarizing the classification of case $m=4$ in chapters \ref{sec:minimumenclosingcircle} and \ref{sec:channelcapacity}, we have the signs of barycentric coordinate and the pair $(Q^\ast,d^\ast)$ as follows:
\begin{align*}
\begin{array}{lll}
& \mbox{\rm Case\,4-1}\ (++++)\ (Q^0,d(P^1,Q^0)) & \\[1mm]
\hline\\[-3mm]
& \mbox{\rm Case\,4-2}\ (-+++) & \\
& \hspace{5mm}\mbox{\rm Case\,4-2{\rm-}1}\ (0+++)\ (Q^1,d(P^2,Q^1)) & \\
& \hspace{5mm}\mbox{\rm Case\,4-2{\rm-}1}\ (0-++)\ (Q^2,d(P^3,Q^2)) & \\[1mm]
\hline\\[-3mm]
& \mbox{\rm Case\,4-3}\ (--++) & \\
& \hspace{5mm}\mbox{\rm Case\,4-3{\rm-}1}\ (0+++)\ (Q^{1(1)},d(P^2,Q^{1(1)})) & \\
& \hspace{5mm}\mbox{\rm Case\,4-3{\rm-}2}\ (0-++)\ (Q^2,d(P^3,Q^2)) & \\
& \hspace{5mm}\mbox{\rm Case\,4-3{\rm-}3}\ (0+-+)\ (Q^2{^\dagger},d(P^2,Q^2{^\dagger})) & 
\end{array}
\end{align*}

In chapters \ref{sec:minimumenclosingcircle} and \ref{sec:channelcapacity}, we used common symbols both in smallest enclosing circle and channel capacity. Then, we will show in TABLE \ref{table:correspondence} the correspondence of symbols between them.
\begin{table}[t]
\caption{Correspondence of symbols between smallest enclosing circle and channel capacity}
\begin{tabular}{l|l|l}
\hline
&
smallest enclosing circle
&
channel capacity
\rule[-1.5mm]{0mm}{5mm}\\
\hline
\hline
\hspace{-1mm}
$\bm\lambda$
&
barycentric coordinate
&
input distribution
\rule[-1.5mm]{0mm}{5mm}\\
\hline
\hspace{-1.5mm}
$P^1,\cdots,P^m$
&
given points $\in\mathbb{R}^n$
&
\hspace{-2mm}\begin{tabular}{l}
output distributions $\in\Delta^n$\hspace*{-11mm}\\
for input symbols\\
$x_1,\cdots,x_m$
\end{tabular}
\rule[-5mm]{0mm}{11.7mm}\\
\hline
\hspace{-1mm}
$\Phi$
&
$\begin{pmatrix}
\ P^1\ \\
\vdots\\
P^m
\end{pmatrix}$
\hspace{-1.5mm}\begin{tabular}{l}
matrix of\\
given points
\end{tabular}
&
$\begin{pmatrix}
\ P^1\ \\
\vdots\\
P^m
\end{pmatrix}$
\hspace{-1.5mm}\begin{tabular}{l}
channel matrix\hspace{-8mm}
\end{tabular}
\rule[-6mm]{0mm}{14mm}\\
\hline
\hspace{-3mm}
\begin{tabular}{l}
convex\\
function
\end{tabular}
&
\hspace{-1.5mm}$\begin{array}{l}
f(\bm\lambda,\Phi)=\\
\hspace{6mm}\bm\lambda{\,^t\hspace{-0.2mm}}\bm a-\bm\lambda\Phi{\,^t\hspace{-0.2mm}}\Phi{\,^t\hspace{-0.3mm}}\bm\lambda\\
\bm a=(\|P^1\|^2,\cdots,\|P^m\|^2)\hspace{-2mm}
\end{array}$
&
\hspace{-1.5mm}$\begin{array}{l}
I({\bm\lambda},\Phi)\\
\text{mutual information}
\end{array}$
\rule[-5mm]{0mm}{11.5mm}\\
\hline
\hspace{-1mm}
metric
&
\hspace{-1.5mm}$\begin{array}{l}
d(P,Q)\\[0.5mm]
\text{Euclidean distance}
\end{array}$
&
\hspace{-1.5mm}$\begin{array}{l}
D(P\|Q)\\[0.5mm]
\text{Kullback-Leibler}\\[-0.5mm]
\text{divergence}
\end{array}$
\rule[-5mm]{0mm}{12mm}\\
\hline
\hspace{-3mm}
\begin{tabular}{l}
Kuhn-\\
Tucker\\
condition
\end{tabular}
& 
\hspace{-1.5mm}$\begin{array}{l}
d(P^i,\bm\lambda^\ast\Phi)\\
\left\{\begin{array}{ll}
\!\!\!=d_0, {\mbox{\rm if}}\ \lambda^\ast_i>0\\
\!\!\!\leq d_0, {\mbox{\rm if}}\ \lambda^\ast_i=0
\end{array}\right.
\end{array}$
& 
\hspace{-1.5mm}$\begin{array}{l}
D(P^i\|\bm\lambda^\ast\Phi)\\
\left\{\begin{array}{ll}
\!\!\!=C_0, {\mbox{\rm if}}\ \lambda^\ast_i>0\\
\!\!\!\leq C_0, {\mbox{\rm if}}\ \lambda^\ast_i=0
\end{array}\right. 
\end{array}$
\rule[-5mm]{0mm}{12mm}\\
\hline
\hspace{-1mm}
$Q^\ast$
&
\hspace{-1.5mm}\begin{tabular}{l}
center of smallest\\
enclosing circle
\end{tabular}
&
\hspace{-1.5mm}\begin{tabular}{l}
capacity achieving\\
output distribution
\end{tabular}
\rule[-3mm]{0mm}{8mm}\\
\hline
\hspace{-3mm}
\begin{tabular}{l}
inner\\
product
\end{tabular}
&
$\ds\sum_{j=1}^n(Q^1_j-Q^2_j)(Q^3_j-Q^2_j)$
&
$\ds\sum_{j=2}^n(\eta^1_j-\eta^2_j)(\theta^3_j-\theta^2_j)$\hspace{-7mm}
\rule[-5mm]{0mm}{10.5mm}\\
\hline
\hspace{-3mm}
\begin{tabular}{l}
Pythagorean\hspace*{-2mm}\\
theorem
\end{tabular}
&
\hspace{-1.5mm}$\begin{array}{l}
d^2(P,Q)+d^2(Q,R)\\
=d^2(P,R)\end{array}$
&
\hspace{-1.5mm}$\begin{array}{l}
D(P\|Q)+D(Q\|R)\\
=D(P\|R)\end{array}$
\rule[-3mm]{0mm}{8mm}\\
\hline
\hspace{-1mm}
projection
&
$\ds\min_{P\in L}d(P,Q)$
&
$\ds\min_{P\in L}D(P\|Q)$
\rule[-3mm]{0mm}{6.5mm}\\
\hline
\end{tabular}
\label{table:correspondence}
\end{table}
\section{Search for optimal solution for arbitrary placement of points}
\label{sec:dimensionlifting}
In the previous chapters, we assumed that the given points $P^1,\cdots,P^m\in\mathbb{R}^n$ or $\in\Delta^n$ are in general position, i.e., (\ref{eqn:rankPsim-1}) or (\ref{eqn:rankA}). Under these assumptions, there exists the equidistant point $Q^0$ from $P^1,\cdots,P^m$, hence by the projection algorithm, we could calculate the center of the smallest enclosing circle $Q^\ast$, or the capacity achieving output distribution $Q^\ast$. However, we cannot expect that arbitrarily given points $P^1,\cdots,P^m$ are in general position, especially if $m$ is large. For example, four points in $\mathbb{R}^2$ are not in general position. If $P^1,\cdots,P^m$ are not in general position, the projection algorithms in the previous chapters cannot be used.

In this chapter, we assume
\begin{align}
\label{eqn:ranklessthanorequalm-1}
\text{rank}\,\Psi\leq m-1\ \,\text{for}\ 
\Psi=\left(\begin{array}{c}
P^2-P^1\\
\ \vdots\\
P^m-P^1
\end{array}\right),
\end{align}
which implies that there is no constraint on the placement of the points $P^1,\cdots,P^m$. Because the projection algorithms in chapters \ref{sec:minimumenclosingcircle}, \ref{sec:channelcapacity} cannot be used in its present form if rank $\Phi<m-1$, we must consider some way to avoid this difficulty. In this chapter, we will consider a method of giving a little deformation to the placement of given points $P^1,\cdots,P^m$ so that the points of the deformed placement are in general position, then apply the projection algorithm. According to the method which will be proposed in this chapter, it is not necessary to check the rank of $\Phi$ in advance. Furthermore, we do not need to know about geometric conditions, such as, a point is contained in the convex hull of several other points, or what the dimension of the subspace spanned by the whole points is, and so on. However, because we make a little deformation to the original problem, there might be some possibility that the obtained result is different from the true solution of the original problem.

\begin{figure}[t]
\begin{center}
\begin{overpic}[width=6cm]{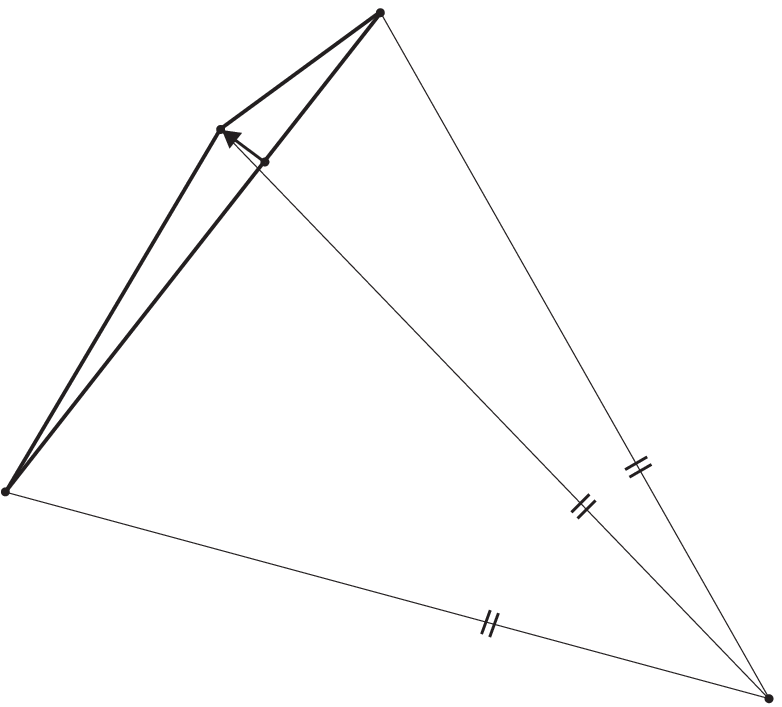}
\put(100,-5){$\widetilde{Q}^0$}
\put(-10,25){$P^2$}
\put(37,68){$P^1$}
\put(18,75){$\widetilde{P}^1$}
\put(52,90){$P^3$}
\end{overpic}
\end{center}
\caption{Getting equidistant point by shifting a point}
\label{fig:5}
\end{figure}

Let us consider an example of Fig.\ref{fig:5}. The three points $P^1,P^2,P^3$ are on a straight line in $\mathbb{R}^2$. In this case, how can we determine the smallest enclosing circle $\Gamma(P^1,P^2,P^3)$ without knowing the positional relationship such as $P^1$ lies between $P^2$ and $P^3$? A method that we conceive immediately is to select two points from the three points and check all the combinations $\Gamma(P^1,P^2)$, $\Gamma(P^1,P^3)$, $\Gamma(P^2,P^3)$. Among them, the one which 
includes all the points with the minimum radius is the smallest enclosing circle. However, if the number of points becomes large the computational complexity becomes large, thus, such a combinatorial method cannot be applied to general cases.

Now, let us shift $P^1$ slightly to have $\widetilde{P}^1$ (see Fig.\,\ref{fig:5}), then there exists the equidistant point $\widetilde{Q}^0$ from $\widetilde{P}^1,P^2,P^3$, so, we can calculate the smallest enclosing circle $\Gamma(\widetilde{P}^1,P^2,P^3)$ by the projection algorithm in the previous chapters. If the amount of shift is small, then we can expect $\Gamma(\widetilde{P}^1,P^2,P^3)=\Gamma(P^1,P^2,P^3)$.

Based on the above fundamental idea, we will give the following algorithm.
\subsection{Method of lifting dimension of point in $\mathbb{R}^n$}
\label{subsec:lifting1}
For $m$ points $P^1,\cdots,P^m$ in $\mathbb{R}^n$ with (\ref{eqn:ranklessthanorequalm-1}), define the points $\widetilde{P}^1,\cdots,\widetilde{P}^m\in\mathbb{R}^{n+m}$ by lifting the dimension as follows:
\begin{align}
\label{eqn:tildePi}
\begin{array}{rl}
P^1=(P^1_1,\cdots,P^1_n)&\hspace{-2mm}\rightarrow\widetilde{P}^1=(P^1_1,\cdots,P^1_n,\varepsilon,0,\cdots,0),\\
P^2=(P^2_1,\cdots,P^2_n)&\hspace{-2mm}\rightarrow\widetilde{P}^2=(P^2_1,\cdots,P^2_n,0,\varepsilon,\cdots,0),\\
\vdots\hspace{10mm}&\hspace{20mm}\vdots\\
P^m=(P^m_1,\cdots,P^m_n)&\hspace{-2mm}\rightarrow\widetilde{P}^m=(P^m_1,\cdots,P^m_n,0,\cdots,0,\varepsilon),
\end{array}
\end{align}
where $\varepsilon\in\mathbb{R},\,\varepsilon\neq0$, and $|\varepsilon|$ is sufficiently small. The above correspondence $P^i\rightarrow\widetilde{P}^i$ can be written as follows. Defining the $i$th fundamental vector $\bm{e}_i\in\mathbb{R}^m$ by
\begin{align}
\bm{e}_i=(0,\cdots,0,\stackrel{i\,\text{th}}{\stackrel{\vee}{1}}\hspace{-1mm},\ 0,\cdots,0),\ i=1,\cdots,m,
\end{align}
we have the correspondence
\begin{align}
\label{eqn:lifting}
\mathbb{R}^n\ni P^i\rightarrow\widetilde{P}^i=(P^i,\varepsilon\bm{e}_i)\in\mathbb{R}^{n+m},\,i=1,\cdots,m.
\end{align}
Then let
\begin{align}
\widetilde\Phi=\left(\begin{array}{c}
\widetilde{P}^1\\
\vdots\\
\widetilde{P}^m
\end{array}
\right)\in\mathbb{R}^{m\times(n+m)}.
\end{align}
We have rank\,$\widetilde\Phi=m$ for $\varepsilon\neq0$, hence, by the same argument as (\ref{sec:quidistantpointanditsbarycentriccoordinate}), we see that there exists the equidistant point $\widetilde{Q}^0\in L(\widetilde{P}^1,\cdots,\widetilde{P}^m)\subset\mathbb{R}^{n+m}$ from $\widetilde{P}^1,\cdots,\widetilde{P}^m$. Denote by $\widetilde{\bm\lambda}^0=(\widetilde{\lambda}^0_1,\cdots,\widetilde{\lambda}^0_m)$ the barycentric coordinate of $\widetilde{Q}^0$ about $\widetilde{P}^1,\cdots,\widetilde{P}^m$. Define $\widetilde{M}\equiv\widetilde\Phi\hspace{0.3mm}{^t}\hspace{-0.3mm}\widetilde\Phi\in\mathbb{R}^{m\times m}$, then since rank\,$\widetilde{M}=$rank\,$\widetilde\Phi=m$, $\widetilde{M}$ is non-singular. Putting $\widetilde{\bm{a}}=(\|\widetilde{P}^1\|^2,\cdots,\|\widetilde{P}^m\|^2)\in\mathbb{R}^m$, 
we have
\begin{align}
\label{eqn:lambdatilde0}
\widetilde{\bm\lambda}^0=\ds\frac{1}{2}\left(\widetilde{\bm{a}}+
\ds\frac{2-\widetilde{\bm{a}}\widetilde{M}^{-1}\hspace{0.3mm}{^t}\hspace{-0.3mm}\bm{1}}{\bm{1}\widetilde{M}^{-1}\hspace{0.3mm}{^t}\hspace{-0.3mm}\bm{1}}
\bm{1}\right)\widetilde{M}^{-1}
\end{align}
in a similar way as (\ref{eqn:barycentric12}).

Suppose we had the center $\widetilde{Q}^\ast$ of the smallest enclosing circle for $\widetilde{P}^1,\cdots,\widetilde{P}^m$ by the projection algorithm in chapter \ref{sec:minimumenclosingcircle}, then for small $\varepsilon$, $\widetilde{Q}^\ast$ is close to the true $Q^\ast$, which is the center of the smallest enclosing circle for $P^1,\cdots,P^m$. Denote by $\widetilde{\bm\lambda}^\ast$ the barycentric coordinate of $\widetilde{Q}^\ast$ about $\widetilde{P}^1,\cdots,\widetilde{P}^m$, and $\widetilde{\bm\lambda}^\ast\big|_{\varepsilon=0}$ by substituting $\varepsilon=0$ in $\widetilde{\bm\lambda}^\ast$. Then $Q^\ast=\widetilde{\bm\lambda}^\ast\big|_{\varepsilon=0}\Phi$ is expected to be the center of the original smallest enclosing circle.
\begin{example}
\label{example:2}
{\rm 
Let us consider three points $P^1=(1),P^2=(0),P^3=(2)$ in $\mathbb{R}$ (see Fig.\ref{fig:6}).
\begin{figure}[t]
\begin{center}
\begin{overpic}[width=7.5cm]{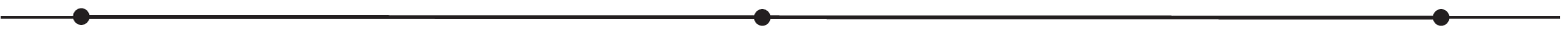}
\put(46,-6){$P^1=(1)$}
\put(2,-6){$P^2=(0)$}
\put(89,-6){$P^3=(2)$}
\end{overpic}
\end{center}
\bigskip
\caption{Three points $P^1=(1),P^2=(0),P^3=(2)$ in $\mathbb{R}$}
\label{fig:6}
\end{figure}
Lifting the dimension, we have $\widetilde{P}^1,\widetilde{P}^2,\widetilde{P}^3\in{\mathbb{R}}^4$ by
\begin{align*}
&P^1=(1)\rightarrow\widetilde{P}^1=(1,\varepsilon,0,0),\\
&P^2=(0)\rightarrow\widetilde{P}^2=(0,0,\varepsilon,0),\\
&P^3=(2)\rightarrow\widetilde{P}^3=(2,0,0,\varepsilon).
\end{align*}
From (\ref{eqn:lambdatilde0}), we have the barycentric coordinate $\widetilde{\bm\lambda}^0=(\widetilde\lambda^0_1,\widetilde\lambda^0_2,\widetilde\lambda^0_3)$ about $\widetilde{P}^1,\widetilde{P}^2,\widetilde{P}^3$ of the equidistant point $\widetilde{Q}^0$ from $\widetilde{P}^1,\widetilde{P}^2,\widetilde{P}^3$ as
\begin{align}
\widetilde{\bm\lambda}^0=\left(
\ds\frac{-2+2\varepsilon^2}{6\varepsilon^2},
\ds\frac{1+2\varepsilon^2}{6\varepsilon^2},
\ds\frac{1+2\varepsilon^2}{6\varepsilon^2}
\right).
\end{align}
For sufficiently small $|\varepsilon|$ with $\varepsilon\neq0$, $\widetilde\lambda^0_1=(-2+2\varepsilon^2)/(6\varepsilon^2)<0$, thus we remove $\widetilde{P}^1$ and project $\widetilde{Q}^0$ onto $L(\widetilde{P}^2,\widetilde{P}^3)$. The barycentric coordinate $\widetilde{\bm\lambda}^1=(\widetilde\lambda^1_1,\widetilde\lambda^1_2,\widetilde\lambda^1_3)$ of $\widetilde{Q}^1=\pi(\widetilde{Q}^0|L(\widetilde{P}^2,\widetilde{P}^3))$ about $\widetilde{P}^1,\widetilde{P}^2,\widetilde{P}^3$ is $\widetilde{\bm\lambda}^1=(0,1/2,1/2)$. Because $\widetilde\lambda^1_i\geq0,i=1,2,3$, this case is the situation 1, so we have $\widetilde{\bm\lambda}^1=\widetilde{\bm\lambda}^\ast$ by Theorem \ref{theorem:2}. Then, substituting $\varepsilon=0$, we obtain $\widetilde{\bm\lambda}^\ast\big|_{\varepsilon=0}=(0,1/2,1/2)$ and 
\begin{align*}
Q^\ast&=\widetilde{\bm\lambda}^\ast\big|_{\varepsilon=0}\Phi\\
&=\left(0,\ds\frac{1}{2},\ds\frac{1}{2}\right)
\left(\begin{array}{c}
1\\
0\\
2
\end{array}\right)\\
&=1.
\end{align*}
Therefore, the center $Q^\ast$ of $\Gamma(P^1,P^2,P^3)$ is obtained correctly.
}
\end{example}
\begin{example}
\label{example:3}
{\rm 
Let us consider four points $P^1=(1,2),P^2=(0,0),$ $P^3=(2,0),P^4=(1,3)$ in $\mathbb{R}^2$, which are not in general position (see Fig.\ref{fig:7}).
\begin{figure}[t]
\begin{center}
\begin{overpic}[width=7cm]{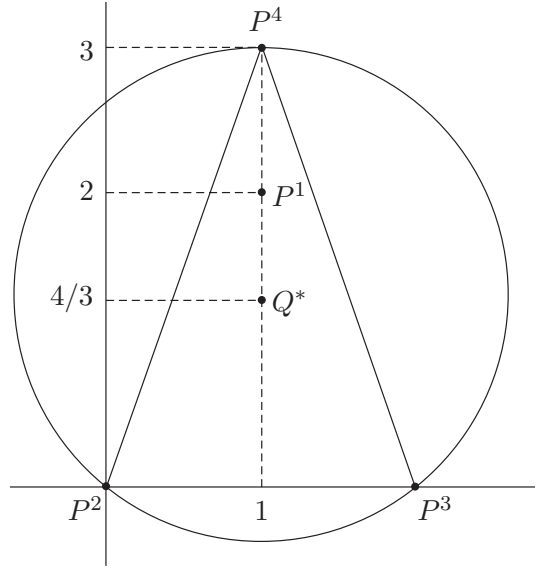}
\put(42,8){$1$}
\put(12,87){$3$}
\put(12,63){$2$}
\put(7,45){$4/3$}
\put(45,62){$P^1$}
\put(10,8){$P^2$}
\put(70,8){$P^3$}
\put(41,92){$P^4$}
\put(45,44){$Q^\ast$}
\end{overpic}
\end{center}
\caption{Four points $P^1=(1,2),P^2=(0,0),P^3=(2,0),P^4=(1,3)$ in $\mathbb{R}^2$}
\label{fig:7}
\end{figure}
By lifting the dimension, we have
\begin{align*}
&P^1=(1,2)\rightarrow\widetilde{P}^1=(1,2,\varepsilon,0,0,0),\\
&P^2=(0,0)\rightarrow\widetilde{P}^2=(0,0,0,\varepsilon,0,0),\\
&P^3=(2,0)\rightarrow\widetilde{P}^3=(2,0,0,0,\varepsilon,0),\\
&P^4=(1,3)\rightarrow\widetilde{P}^4=(1,3,0,0,0,\varepsilon).
\end{align*}
Let $\widetilde{Q}^0$ be the equidistant point from $\widetilde{P}^1,\cdots,\widetilde{P}^4$, and $\widetilde{\bm\lambda}^0=(\widetilde\lambda^0_1,\cdots,\widetilde\lambda^0_4)$ be the barycentric coordinate of $\widetilde{Q}^0$ about $\widetilde{P}^1,\cdots,\widetilde{P}^4$. From (\ref{eqn:lambdatilde0}),
%
%
%
\begin{align}
\widetilde{\bm\lambda}^0=
\left(\ds\frac{-42+7\varepsilon^2+2\varepsilon^4}{54\varepsilon^2+8\varepsilon^4}, 
\ds\frac{7+15\varepsilon^2+2\varepsilon^4}{54\varepsilon^2+8\varepsilon^4}, 
\ds\frac{7+15\varepsilon^2+2\varepsilon^4}{54\varepsilon^2+8\varepsilon^4}, 
\ds\frac{28+17\varepsilon^2+2\varepsilon^4}{54\varepsilon^2+8\varepsilon^4}\right).
\end{align}
For sufficiently small $|\varepsilon|$ with $\varepsilon\neq0$, $\widetilde\lambda^0_1<0$, thus we remove $\widetilde{P}^1$ and project $\widetilde{Q}^0$ onto $L(\widetilde{P}^2,\widetilde{P}^3,\widetilde{P}^4)$. The barycentric coordinate $\widetilde{\bm\lambda}^1=(\widetilde\lambda^1_1,\cdots,\widetilde\lambda^1_4)$ about $\widetilde{P}^1,\cdots,\widetilde{P}^4$ of $\widetilde{Q}^1=\pi(\widetilde{Q}^0|L(\widetilde{P}^2,\widetilde{P}^3,\widetilde{P}^4))$ is
\begin{align}
\widetilde{\bm\lambda}^1=\left(0,\ds\frac{5+\varepsilon^2}{18+3\varepsilon^2}, 
\ds\frac{5+\varepsilon^2}{18+3\varepsilon^2}, 
\ds\frac{8+\varepsilon^2}{18+3\varepsilon^2}\right).
\end{align}
Because $\widetilde\lambda^1_i\geq0,\,i=1,\cdots4$, this case is the situation 1, so we have $\widetilde{\bm\lambda}^1=\widetilde{\bm\lambda}^\ast$ by Theorem \ref{theorem:2}. Substituting $\varepsilon=0$, we have $\widetilde{\bm\lambda}^\ast\big|_{\varepsilon=0}=(0,5/18,5/18,8/18)$. The center $Q^\ast$ of $\Gamma(P^1,\cdots,P^4)$ is 

\begin{align}
Q^\ast&=\widetilde{\bm\lambda}^\ast\big|_{\varepsilon=0}\Phi\\
&=\left(0,\ds\frac{5}{18},\ds\frac{5}{18},\ds\frac{8}{18}\right)
\left(\begin{array}{cc}
1 & 2\\
0 & 0\\
2 & 0\\
1 & 3
\end{array}\right)\\
&=\left(1,\,\ds\frac{4}{3}\right).
\end{align}
Hence, also in this case the center $Q^\ast$ of $\Gamma(P^1,\cdots,P^4)$ is correctly obtained.
}
\end{example}
\subsection{Lifting dimension of channel matrix}
In the previous section, for the points $P^1,\cdots,P^m$ in $\mathbb{R}^n$, by lifting the dimension of these points, we had $\widetilde{P}^1,\cdots,\widetilde{P}^m\in\mathbb{R}^{n+m}$ which are in general position. In this section, for the row vectors $P^1,\cdots,P^m$ of the channel matrix $\Phi$ in (\ref{eqn:thechannelmatrix}), we will define $\widetilde{P}^1,\cdots,\widetilde{P}^m$ by lifting the dimension so that they are in general position.

In the case of smallest enclosing circle, we added $\varepsilon\bm{e}_i$ to $P^i$ for lifting the dimension in (\ref{eqn:lifting}). $\varepsilon\bm{e}_i$ is in the vicinity of the origin of $\mathbb{R}^n$. In the problem of channel capacity, we will add a constant multiple of a distribution in the vicinity of the uniform distribution to a constant multiple of the row vector $P^i$ of the channel matrix. Since the sum of the probabilities is 1, if some component is increased then another component must be decreased. By considering so, we will use a vector $\left((1+\varepsilon)/2m,(1-\varepsilon)/2m,1/2m,\cdots,1/2m)\right)$ etc., which is in the vicinity of the uniform distribution $(1/2m,1/2m,\cdots,1/2m)\in\Delta^{2m}$. Now, define $\widetilde{P}^i\in\Delta^{n+2m}$ by
%
%
%
\begin{align}
\widetilde{P}^i=\Big(\ds\frac{P^i_1}{2m+1},\cdots,\ds\frac{P^i_n}{2m+1}, 
\overbrace{\frac{1}{2m+1},\cdots,\frac{1}{2m+1}}^{2i-2}, 
\stackrel{2i-1\,\text{th}}{\stackrel{\vee}{\frac{1+\epsilon}{2m+1}}}, 
\stackrel{2i\,\text{th}}{\stackrel{\vee}{\frac{1-\epsilon}{2m+1}}}, 
\overbrace{\frac{1}{2m+1},\cdots,\frac{1}{2m+1}}^{2m-2i}\Big),\label{eqn:phat}
\end{align}
where $|\varepsilon|$ is sufficiently small and $\varepsilon\neq0$. Let
\begin{align}
\label{eqn:Phitilde}
\widetilde\Phi=\left(\begin{array}{c}\widetilde{P}^1\\\vdots\\\widetilde{P}^m\end{array}\right),
\end{align}
then rank\,$\widetilde\Phi=m$ for $\varepsilon\neq0$.

Define an $m\times2m$ matrix $\Xi$ by
%
%
%
\begin{align*}
\Xi=\frac{1}{2m}
\left(\begin{array}{ccccccc}
1+\epsilon & 1-\epsilon & 1 & 1 & \cdots & 1 & 1\\
1 & 1 & 1+\epsilon & 1-\epsilon & \cdots & 1 & 1\\
\vdots & \vdots && \ddots & \ddots& \vdots & \vdots\\
1 & 1 & 1 & 1 & \cdots & 1+\epsilon & 1-\epsilon\\
\end{array}\right),
\end{align*}
and denote by $\Xi^i$ the $i$ th row vector of $\Xi$, where $\Xi^i\in\Delta^{2m},\,i=1,\cdots,m$. (\ref{eqn:phat}) can be written as
\begin{align}
\label{:eqn:piti}
\widetilde{P}^i=\left(\frac{1}{2m+1}P^i,\frac{2m}{2m+1}\Xi^i\right),\ i=1,\cdots,m.
\end{align}
\subsubsection{Dummy output alphabet}
The channel defined by (\ref{eqn:phat}), (\ref{eqn:Phitilde}),\,(\ref{:eqn:piti}) is modeled as follows. That is, let the input alphabet $\{x_1,\cdots,x_m\}$ be the same as the original one and then to the original output alphabet $\{y_1,$ $\cdots,y_n\}$ dummy alphabet $\{y'_1,\cdots,y'_{2m}\}$ is newly added. Let us define random variables $Y$, $Y'$ taking values on $\{y_1,\cdots,y_n\}$, $\{y'_1,\cdots,y'_{2m}\}$, respectively, by
\begin{align*}
&P(Y=y_j|X=x_i)=P^i_j,\,i=1,\cdots,m,j=1,\cdots,n,\\
&P(Y'=y'_{j'}|X=x_i)=\Xi^i_{j'},\,i=1,\cdots,m,j'=1,\cdots,2m,
\end{align*}
where $\Xi^i_{j'}$ denote the $j'$th element of the row vector $\Xi^i$. Further, define a random variable $\widetilde{Y}$ as the mixture of $Y$ and $Y'$ with weight $(1/(2m+1),2m/(2m+1))$. Then $X\rightarrow\widetilde{Y}$ is a model of the channel defined by (\ref{eqn:phat}),\,(\ref{eqn:Phitilde}), (\ref{:eqn:piti}). The mutual information $I(\bm\lambda,\Xi)$ between $X$ and $Y'$ is close to $0$. Dummy alphabet is used for lifting the dimension of the row vectors of the channel matrix to make them be in general position.
\subsubsection{Equation of barycentric coordinate}
Now, for $Q=(Q_1,\cdots,Q_n)\in\Delta^n$, $T=(T_1,\cdots,T_{2m})\in\Delta^{2m}$, let $\widetilde{Q}\in\Delta^{n+2m}$ be
\begin{align}
\label{eqn:qcordinate}
\widetilde{Q}=\left(\frac{1}{2m+1}Q,\ \frac{2m}{2m+1}T\right).
\end{align}
We will calculate the equidistant point $\widetilde{Q}\in L(\widetilde{P}^1\cdots,\widetilde{P}^m)\subset\Delta^{n+2m}$ from $\widetilde{P}^1\cdots,\widetilde{P}^m$, i.e., $\widetilde{Q}$ satisfies 
\begin{align}
\label{eqn:divergence1}
D(\widetilde{P}^i||\widetilde{Q})=D(\widetilde{P}^1||\widetilde{Q}),\ i=2,\cdots,m.
\end{align}
By calculation,
%
%
%
\begin{align}
\label{eqn:divergence2}
D(\widetilde{P}^i||\widetilde{Q})=
\ds\frac{-1}{2m+1}\Big\{\ds\sum_{j=1}^nP^i_j\log Q_j
+\epsilon\log\ds\frac{T_{2i-1}}{T_{2i}}+H(P^i)
+\sum_{j'=1}^{2m}\log(2mT_{j'})+\bar{H}(\epsilon)\Big\},
\end{align}
where we set $H(P^i)=-\sum_{j=1}^nP^i_j\log P^i_j$, $\bar{H}(\epsilon)=-(1+\epsilon)\log(1+\epsilon)-(1-\epsilon)\log(1-\epsilon)$. By (\ref{eqn:divergence1}), (\ref{eqn:divergence2}), we have
%
%
%
\begin{align}
\label{eqn:divergence3}
\sum_{j=1}^n\left(P^i_j-P^1_j\right)\log Q_j+\epsilon\log\ds\frac{T_{2i-1}}{T_{2i}}
-\epsilon\log\ds\frac{T_1}{T_2}
=-H(P^i)+H(P^1),\,i=2,\cdots,m.
\end{align}
Next, let $\widetilde{\bm\lambda}=(\widetilde\lambda_1,\cdots,\widetilde\lambda_m)$ be the barycentric coordinate of $\widetilde{Q}$ about $\widetilde{P}^1,\cdots,\widetilde{P}^m$, i.e., $\widetilde{Q}=\widetilde{\bm\lambda}\widetilde\Phi$ from (\ref{eqn:Phitilde}). By (\ref{:eqn:piti}), (\ref{eqn:qcordinate}), $Q=\widetilde{\bm\lambda}\Phi$, $T=\widetilde{\bm\lambda}\,\Xi$, thus substituting these into (\ref{eqn:divergence3}), we have
%
%
%
\begin{align}
\label{eqn:divergence4}
\sum_{j=1}^n\left(P^i_j-P^1_j\right)\log\sum_{i'=1}^m\widetilde\lambda_{i'}P^{i'}_j+
\epsilon\log\frac{1+\varepsilon\widetilde\lambda_i}{1-\varepsilon\widetilde\lambda_i}
-\epsilon\log\frac{1+\varepsilon\widetilde\lambda_1}{1-\varepsilon\widetilde\lambda_1}
=-H(P^i)+H(P^1),\ i=2,\cdots,m.
\end{align}
For $m$ unknowns $\widetilde\lambda_1,\cdots,\widetilde\lambda_m$, there are $m-1$ equations (\ref{eqn:divergence4}) and one equation $\sum_{i=1}^m\widetilde\lambda_i=1$, then we have the solution $\widetilde{\bm\lambda}$. The existence and the uniqueness of the solution $\widetilde{\bm\lambda}$ is guaranteed by Lemma \ref{lem:25}.
\begin{example}
\label{example:4}
{\rm 
Let us consider a channel matrix
\begin{align}
\Phi=\left(\begin{array}{c}P^1\\ P^2\\ P^3\end{array}\right)
=\left(\begin{array}{cc}0.1 & 0.9\\0.7 & 0.3 \\0.8 & 0.2\end{array}\right),
\end{align}
where the rows are not in general position (see Fig.\ref{fig:8}). By lifting the dimension of $\Phi$, we have the channel matrix $\widetilde\Phi$ of (\ref{eqn:Phitilde}) by
\begin{align}
&P^1=(0.1,0.9)\rightarrow\widetilde{P}^1=(0.1,0.9,1+\varepsilon,1-\varepsilon,1,1,1,1)/7,\nonumber\\
&P^2=(0.7,0.3)\rightarrow\widetilde{P}^2=(0.7,0.3,1,1,1+\varepsilon,1-\varepsilon,1,1)/7,\nonumber\\
&P^3=(0.8,0.2)\rightarrow\widetilde{P}^3=(0.8,0.2,1,1,1,1,1+\varepsilon,1-\varepsilon)/7.\nonumber
\end{align}
Let $\widetilde{\bm\lambda}^0$ be the barycentric coordinate about $\widetilde{P}^1,\widetilde{P}^2,\widetilde{P}^3$ of the equidistant point $\widetilde{Q}^0$ from $\widetilde{P}^1,\widetilde{P}^2,\widetilde{P}^3$. Solving the equation $(\ref{eqn:divergence4})$ with $\varepsilon=0.05$, we have
\begin{align}
\widetilde{\bm\lambda}^0=(2.39,\,-12.96,\,11.57).
\end{align}
Since $\widetilde\lambda^0_2=-12.96<0$, we remove $\widetilde{P}^2$ and project $\widetilde{Q}^0$ onto $L(\widetilde{P}^1,\widetilde{P}^3)$ to have $\widetilde{Q}^1=\pi(\widetilde{Q}^0|L(\widetilde{P}^1,\widetilde{P}^3))$. Let $\widetilde{\bm\lambda}^1$ be the barycentric coordinate of $\widetilde{Q}^1$ about $\widetilde{P}^1,\widetilde{P}^2,\widetilde{P}^3$. By calculation $\widetilde{\bm\lambda}^1=(0.52,0,0.48)$, so this case is the situation 1. Therefore, we have $\widetilde{\bm\lambda}^\ast=\widetilde{\bm\lambda}^1$ by Theorem \ref{theorem:13} and $\widetilde{\bm\lambda}^\ast\big|_{\varepsilon=0}=(0.52,0,0.48)$. Thus, 
\begin{align}
Q^\ast&=\widetilde{\bm\lambda}^\ast\big|_{\varepsilon=0}\Phi\\
&=(0.52,0,0.48)
\left(\begin{array}{cc}
0.1 & 0.9\\
0.7 & 0.3\\
0.8 & 0.2
\end{array}\right)\\
&=\left(0.436,\,0.564\right),
\end{align}
(see Fig.\ref{fig:8}). 
The channel capacity is $C=D(P^1\|Q^\ast)=0.398$ [bit/symbol]. So, the capacity achieving $Q^\ast$ and the capacity $C$ of $\Phi$ are correctly obtained.
\begin{figure}[t]
\begin{center}
\begin{overpic}[width=7cm]{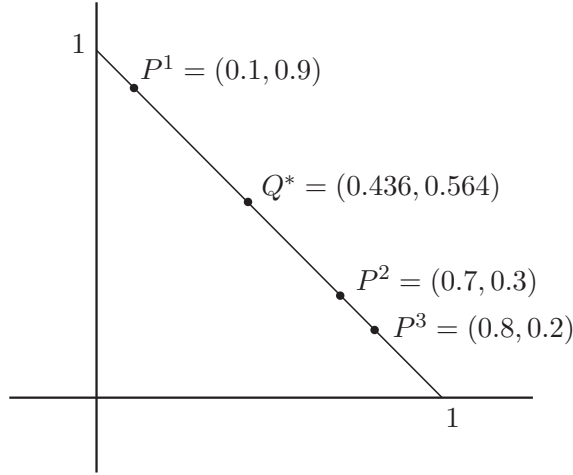}
\put(25,75){$P^1=(0.1,0.9)$}
\put(48,53){$Q^\ast=(0.436,0.564)$}
\put(66,35){$P^2=(0.7,0.3)$}
\put(73,26){$P^3=(0.8,0.2)$}
\put(83,9){$1$}
\put(12,80){$1$}
\end{overpic}
\end{center}
\caption{Three points $P^1=(0.1,0.9),P^2=(0.7,0.3),P^3=(0.8,0.2)$ in $\Delta^2$}
\label{fig:8}
\end{figure}
}
\end{example}
\begin{example}
\label{example:5}
{\rm 
Consider the channel matrix 
\begin{align}
\Phi=\left(\begin{array}{c}P^1\\ P^2\\ P^3\\ P^4\end{array}\right)
=\left(\begin{array}{ccc}
2/5 & 2/5 & 1/5\\
1/3 & 1/3 & 1/3\\
4/5 & 1/10 & 1/10\\
1/10 & 4/5 & 1/10
\end{array}\right),
\end{align}
where the rows are not in general position (see Fig.\ref{fig:9}). By lifting the dimension of $\Phi$, we have the channel matrix $\widetilde\Phi$ of (\ref{eqn:Phitilde}) by
%
%
%
\begin{align*}
P^1=(2/5,2/5,1/5)
&\rightarrow\widetilde{P}^1=(2/5,2/5,1/5,1+\varepsilon,1-\varepsilon,1,1,1,1,1,1)/9,\\
P^2=(1/3,1/3,1/3)
&\rightarrow\widetilde{P}^2=(1/3,1/3,1/3,1,1,1+\varepsilon,1-\varepsilon,1,1,1,1)/9,\\
P^3=(4/5,1/10,1/10)
&\rightarrow\widetilde{P}^3=(4/5,1/10,1/10,1,1,1,1,1+\varepsilon,1-\varepsilon,1,1)/9,\\
P^4=(1/10,4/5,1/10)
&\rightarrow\widetilde{P}^4=(1/10,4/5,1/10,1,1,1,1,1,1,1+\varepsilon,1-\varepsilon)/9.
\end{align*}
Solving the equation $(\ref{eqn:divergence4})$ with $\varepsilon=0.05$, we have the barycentric coordinate $\widetilde{\bm\lambda}^0$ of the equidistant point $\widetilde{Q}^0$ from $\widetilde{P}^1,\widetilde{P}^2,\widetilde{P}^3,\widetilde{P}^4$ as 
\begin{align}
\widetilde{\bm\lambda}^0=(-19.00, 7.98, 6.01, 6.01).
\end{align}
Since $\widetilde\lambda^0_1=-19.00<0$, we remove $\widetilde{P}^1$ and consider the projection $\widetilde{Q}^1=\pi(\widetilde{Q}^0|L(\widetilde{P}^2,\widetilde{P}^3,\widetilde{P}^4))$. We have the barycentric coordinate of $\widetilde{Q}^1$ as
\begin{align}
\widetilde{\bm\lambda}^1=(0,-0.14, 0.57, 0.57).
\end{align}
Since $\widetilde{\bm\lambda}^1_2=-0.14<0$, we further we remove $\widetilde{P}^2$ and consider the projection $\widetilde{Q}^2=\pi(\widetilde{Q}^1|L(\widetilde{P}^3,\widetilde{P}^4))$. We have the barycentric coordinate of $\widetilde{Q}^2$ as
\begin{align}
\widetilde{\bm\lambda}^2=(0,0,1/2,1/2).
\end{align}
Thus, this case is the situation 1, so we have $\widetilde{\bm\lambda}^\ast=\widetilde{\bm\lambda}^2$ from Theorem \ref{theorem:13}. Therefore,
\begin{align}
Q^\ast&=\widetilde{\bm\lambda}^\ast\big|_{\varepsilon=0}\Phi\\
&=(0,0,1/2,1/2)
\left(\begin{array}{ccc}
2/5 & 2/5 & 1/5\\
1/3 & 1/3 & 1/3\\
4/5 & 1/10 & 1/10\\
1/10 & 4/5 & 1/10
\end{array}\right)\\
&=\left(9/20,9/20,1/19\right),
\end{align}
and the channel capacity is $C=D(P^3\|Q^\ast)=0.540$ [bit/symbol]. Also in this case, the capacity achieving $Q^\ast$ and the capacity $C$ of $\Phi$ are correctly obtained.
\begin{figure}[t]
\begin{center}
\begin{overpic}[width=6.5cm]{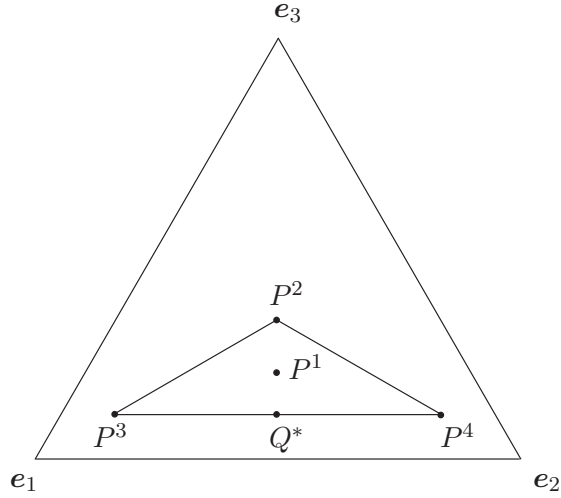}
\put(52,17){$P^1$}
\put(48,32){$P^2$}
\put(12,3){$P^3$}
\put(83,3){$P^4$}
\put(48,3){$Q^\ast$}
\put(-5,-5){$\bm{e}_1$}
\put(102,-5){$\bm{e}_2$}
\put(49,91){$\bm{e}_3$}
\end{overpic}
\end{center}
\bigskip
\caption{Four points $P^1=(2/5,2/5,1/5),P^2=(1/3,1/3,1/3),P^3=(4/5,1/10,1/10),P^4=(1/10,4/5,1/10)$ in $\Delta^3$}
\label{fig:9}
\end{figure}
}
\end{example}

\section{Heuristic algorithm of calculating smallest enclosing circle and channel capacity}
In this chapter, we will propose heuristic projection algorithms for the calculation of the smallest enclosing circle and the channel capacity with arbitrary $m$, $n$ and arbitrary placement of points, based on the algorithms in chapter \ref{sec:minimumenclosingcircle}, \ref{sec:channelcapacity} and the dimension lifting in chapter \ref{sec:dimensionlifting}. In chapter \ref{sec:minimumenclosingcircle}, \ref{sec:channelcapacity}, we gave algorithms to calculate the smallest enclosing circle and the channel capacity under the limited situations, i.e., situation 1 and situation 2. The methods which will be proposed in this chapter can be applied to any number of points and any placement of points, but the obtained results are not always correct. In the following, we describe heuristic algorithms and apply them to many concrete problems generated by random numbers, then show the percentage of getting correct solutions.
\subsection{Heuristic algorithm for Smallest enclosing circle (HS)}
\begin{enumerate}
\item For $m$ points $P^1,\cdots,P^m$ in $\mathbb{R}^n$, lift the dimension by (\ref{eqn:tildePi}) or (\ref{eqn:lifting}) to obtain $\widetilde{P}^1,\cdots,\widetilde{P}^m$.
\item Calculate by (\ref{eqn:lambdatilde0}) the barycentric coordinate $\widetilde{\bm\lambda}^0=(\widetilde\lambda^0_1,\cdots,\widetilde\lambda^0_m)$ of the equidistant point from $\widetilde{P}^1,\cdots,\widetilde{P}^m$.
\item If $\widetilde\lambda^0_i\geq0,\,i=1,\cdots,m$, then end the algorithm and output $\widetilde{Q}^0$.
\item If some of $\widetilde\lambda^0_i$ are negative, consider the smallest one, i.e., the negative one with maximum absolute value, say, it is supposed to be $\widetilde\lambda^0_1$. Then remove $\widetilde{P}^1$ and leave $\widetilde{P}^2,\dots,\widetilde{P}^m$.
\item For $m-1$ points $\widetilde{P}^2,\dots,\widetilde{P}^m$, repeat the algorithm from 2).
\end{enumerate}
The simulation results of the above algorithm are shown in TABLE \ref{table:HS}. Simulation method is as follows: Generate $n$ uniform random integers from $-1000$ to $1000$ and let them be the coordinate of one point in $\mathbb{R}^n$. Repeat it $m$ times to obtain $m$ points of $\mathbb{R}^n$, and let them $P^1,\cdots,P^m$. Calculate the center $Q^\ast$ of the smallest enclosing circle by the heuristic algorithm HS. The above $m$ time generation of points and the calculation of $Q^\ast$ are defined to be one set. Then 10000 sets are executed for each pair of $m,n$ in TABLE \ref{table:HS} to show the percentage that the correct $Q^\ast$ is obtained. Hence, in total, 20 million sets are executed in the whole TABLE \ref{table:HS}.

In every case of TABLE \ref{table:HS}, we have succeeded in obtaining the correct solutions at fairly high percentages. From these results, we can consider that many of actual placements of points are the situation 1. For a fixed value of $m$, the percentage of success increases as $n$ increases. This is because the placements of points are easier to become in general position for a larger dimension $n$. Further, for a fixed value of $n$, the percentage of success decreases as $m$ increases. This is because  there exist so many points in a low dimensional space that our HS becomes difficult to succeed.
\begin{table}[t]
\caption{Simulation results of heuristic algorithm for smallest enclosing circle (HS)}
\begin{center}
$\begin{array}{|c|r|r|r|r|}
\hline
$\backslashbox{\hspace*{3mm}$m$}{\vspace*{-2mm}$n$}$ & 2\ \ \  & 3\ \ \  & 10\ \  & 20\ \ \rule[0mm]{0mm}{0mm}\\
\hline
3 & 100\% & 100\% & 100\% & 100\%\rule[-1.5mm]{0mm}{5mm}\\
\hline
4 & 97.6\% & 99.7\% & 100\% & 100\%\rule[-1.5mm]{0mm}{5mm}\\
\hline
5 & 96.3\% & 95.0\% & 100\% & 100\%\rule[-1.5mm]{0mm}{5mm}\\
\hline
8 & 94.4\% & 85.3\% & 99.5\% & 100\%\rule[-1.5mm]{0mm}{5mm}\\
\hline
10 & 92.9\% & 82.5\% & 98.5\% &100\%\rule[-1.5mm]{0mm}{5mm}\\
\hline
\end{array}$
\end{center}
\label{table:HS}
\end{table}
\begin{figure}[t]
\begin{center}
\begin{overpic}[width=7.5cm]{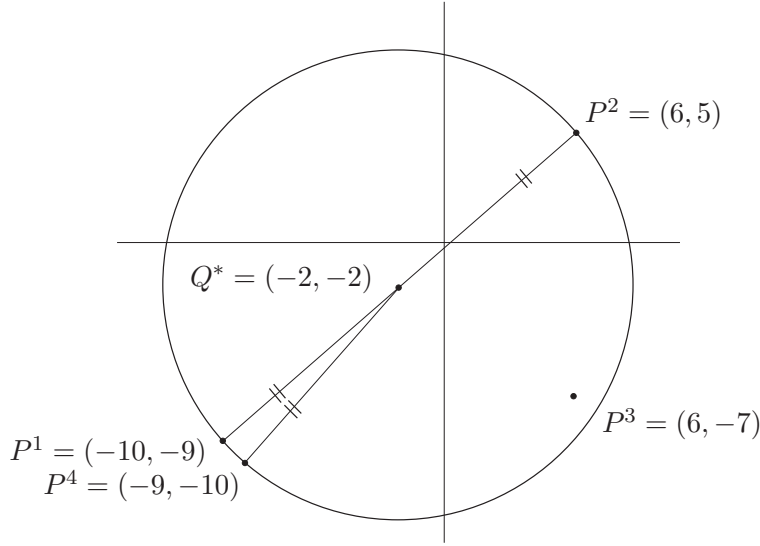}
\put(-19,15){$P^1=(-10,-9)$}
\put(83,75){$P^2=(6,5)$}
\put(86,20){$P^3=(6,-7)$}
\put(-13,9){$P^4=(-9,-10)$}
\put(13,46){$Q^\ast=(-2,-2)$}
\end{overpic}
\end{center}
\caption{Example of placement of points for which our algorithm HS fails}
\label{fig:10}
\end{figure}
\subsubsection{Example of placement of points for which our algorithm fails}
We will show an example of the placement of points for which the smallest enclosing circle cannot be obtained by our heuristic algorithm HS (see Fig.\ref{fig:10}). Consider 4 points $P^1=(-10,-9), P^2=(6,5), P^3=(6,-7), P^4=(-9,-10)$ in $\mathbb{R}^2$. The center $Q^\ast$ of $\Gamma=\Gamma(P^1,\cdots,P^4)$ is $Q^\ast=(-2,-2)$, which is the midpoint of $P^1P^2$. Thus, $P^1$ is necessary for determining $\Gamma$. However, if we apply our HS to this placement of points, we have
%
%
%
\begin{align*}
\bm\lambda^0=&\left(\ds\frac{-64800+22536\varepsilon^2+357\varepsilon^4+\varepsilon^6}{70488\varepsilon^2+1542\varepsilon^4+4\varepsilon^6},\right.
\left.\ds\frac{6480+43350\varepsilon^2+523\varepsilon^4+\varepsilon^6}{70488\varepsilon^2+1542\varepsilon^4+4\varepsilon^6},\right.\\[1mm]
&\hspace{4mm}\left.\ds\frac{-10800-13278\varepsilon^2+319\varepsilon^4+\varepsilon^6}{70488\varepsilon^2+1542\varepsilon^4+4\varepsilon^6},\right.
\left.\ds\frac{69120+17880\varepsilon^2+343\varepsilon^4+\varepsilon^6}{70488\varepsilon^2+1542\varepsilon^4+4\varepsilon^6}\right).
\end{align*}
Because $\lambda^0_1$ is negative and the smallest among $\lambda^0_1,\cdots,\lambda^0_4$ for sufficiently small $|\varepsilon|$ with $\varepsilon\neq0$, our HS removes $P^1$. Hence, in this case our HS fails.
\subsection{Heuristic algorithm for Channel capacity (HC)}
\begin{enumerate}
\item For $m$ probability distributions $P^1,\cdots,P^m$ in $\Delta^n$, lift the dimension by (\ref{eqn:phat}) or (\ref{:eqn:piti}) to obtain $\widetilde{P}^1,\cdots,\widetilde{P}^m$.
\item Calculate by (\ref{eqn:divergence4}) the barycentric coordinate $\widetilde{\bm\lambda}^0=(\widetilde\lambda^0_1,\cdots,\widetilde\lambda^0_m)$ of the equidistant point from $\widetilde{P}^1,\cdots,\widetilde{P}^m$.
\item If $\widetilde\lambda^0_i\geq0,\,i=1,\cdots,m$, then end the algorithm and output $\widetilde{Q}^0$.
\item If some of $\widetilde\lambda^0_i$ are negative, consider the smallest one, i.e., the negative one with maximum absolute value, say, it is supposed to be $\widetilde\lambda^0_1$. Then remove $\widetilde{P}^1$ and leave $\widetilde{P}^2,\dots,\widetilde{P}^m$.
\item For $m-1$ points $\widetilde{P}^2,\dots,\widetilde{P}^m$, repeat the algorithm from 2).
\end{enumerate}
The simulation results of the above algorithm are shown in TABLE \ref{table:HC}. Simulation method is as follows: Generate $n$ uniform random numbers $U_j$ in $(0,1),\ j=1,\cdots,n$ and put $S=\sum_{j=1}^nU_j$. Then define $P=(U_1/S,\cdots,U_n/S)\in\Delta^n$. Repeat it $m$ times to obtain $m$ probability distributions $P^1,\cdots,P^m\in\Delta^n$. Calculate the capacity achieving $Q^\ast$ by the heuristic algorithm HC. The above $m$ time generation of probability distributions and the calculation of $Q^\ast$ are defined to be one set. Then 10000 sets are executed for each pair of $m,n$ in TABLE \ref{table:HC} to show the percentage that the correct $Q^\ast$ is obtained. The failures of HC in TABLE \ref{table:HC} include computational errors by the function FindRoot in Mathematica.

In every case of TABLE \ref{table:HC}, the percentage of success is more than 99\%. But, in comparison with TABLE \ref{table:HS} of the smallest enclosing circle, we do not understand well the reason why TABLE \ref{table:HC} has higher rate of success, although the geometric structures of the two problems are similar.
\begin{table}[t]
\caption{Simulation results of heuristic algorithm for channel capacity (HC)}
\begin{center}
$\begin{array}{|c|r|r|r|r|}
\hline
$\backslashbox{\hspace*{3mm}$m$}{\vspace*{-2mm}$n$}$ & 2\ \ \  & 3\ \ \  & 10\ \  & 20\ \ \rule[0mm]{0mm}{0mm}\\
\hline
3 & 99.9\% & 99.9\% & 99.9\% & 99.9\%\rule[-1.5mm]{0mm}{5mm}\\
\hline
4 & 99.9\% & 99.9\% & 99.9\% & 99.9\%\rule[-1.5mm]{0mm}{5mm}\\
\hline
5 & 99.9\% & 99.8\% & 99.9\% & 99.9\%\rule[-1.5mm]{0mm}{5mm}\\
\hline
8 & 99.9\% & 99.8\% & 99.9\% & 99.9\%\rule[-1.5mm]{0mm}{5mm}\\
\hline
10 & 99.8\% & 99.5\% & 99.8\% & 99.9\%\rule[-1.5mm]{0mm}{5mm}\\
\hline
\end{array}$
\end{center}
\label{table:HC}
\end{table}
\subsection{Comparison of computational complexity}
For given $m$ points $P^1,\cdots,P^m\in\mathbb{R}^n$ or $\in\Delta^n$, the simplest way to calculate the center of the smallest enclosing circle or the channel capacity achieving distribution is to examine all the possible combinations of the points \cite{mur}. This is a brute force method. For example, consider the case of 4 points in $\mathbb{R}^2$. Because a circle in $\mathbb{R}^2$ is determined by two or three points, the total number of all the possible combinations of the points in the brute force method is ${_4}C_2+{_4}C_3=10$. In a similar way, the total number $N_1$ of all the possible combinations of $m$ points in $\mathbb{R}^n$ or $\Delta^n$ is
\begin{align}
N_1=\ds\sum_{\ell=0}^{\min(m,n+1)}{_m}C_\ell.
\end{align}
If we consider the computational complexity for one combination as 1, then the computational complexity of the brute force method is $N_1$.

On the other hand, in our proposed heuristic algorithms, the number of points is decreasing by one at every time, hence the computational complexity $N_2$ is
\begin{align}
N_2=m-2.
\end{align}
We will show in TABLE \ref{table:complexity} the ratio $N_2/N_1$ for the case of TABLE \ref{table:HS} and \ref{table:HC}. $N_2$ is very smaller than $N_1$, however, note that the proposed method does not necessarily yield the correct solutions.
\begin{table}[t]
\caption{Ratio $N_2/N_1$ of computational complexity}
\begin{center}
$\begin{array}{|c|r|r|r|r|}
\hline
$\backslashbox{\hspace*{3mm}$m$}{\vspace*{-2mm}$n$}$ & 2\ \ \  & 3\ \ \  & 10\ \  & 20\ \ \rule[0mm]{0mm}{0mm}\\
\hline
3 & 0.250 & 0.250 & 0.250 & 0.250\rule[-1.5mm]{0mm}{5mm}\\
\hline
4 & 0.200 & 0.182 & 0.182 & 0.182\rule[-1.5mm]{0mm}{5mm}\\
\hline
5 & 0.150 & 0.120 & 0.115 & 0.115\rule[-1.5mm]{0mm}{5mm}\\
\hline
8 & 0.071 & 0.039 & 0.024 & 0.024\rule[-1.5mm]{0mm}{5mm}\\
\hline
10 & 0.048 & 0.021 & 0.008 & 0.008\rule[-1.5mm]{0mm}{5mm}\\
\hline
\end{array}$
\end{center}
\label{table:complexity}
\end{table}
\section{Conclusion}
Since the Euclidean geometry is familiar to us, it is easy to develop an algorithm and to prove the correctness of the obtained algorithm. A proposition in the Euclidean geometry can be proved in many ways because there are many tools that we can use. However, if we consider the corresponding proposition in the information geometry, all the proofs in the Euclidean geometry are not necessarily applicable to the proposition. We found that there is one natural proof in the Euclidean geometry that is directly applicable to the corresponding proposition of the information geometry. Such a proof uses only common properties of both Euclidean and information geometries. We also found that the natural proof is the simplest proof. This is a very interesting result. If we considered only the problem of channel capacity from the beginning, I think we could not obtain the projection algorithm developed in this paper. It is a result of a successful link between the Euclidean and information geometries.

\section{Future works}
\begin{itemize}
\item[(1)] Make an algorithm to calculate $Q^\ast$ for arbitrary number $m$ of points in general position.
\item[(2)] Make a projection algorithm for the rate distortion function and the capacity constraint function.
\item[(3)] Transplant the Arimoto algorithm to the problem of smallest enclosing circle.
\end{itemize}

\end{document}